\documentclass[showpacs,preprintnumbers,amsmath,amssymb,prd]{revtex4}

\usepackage{graphicx}
\usepackage{dcolumn}
\usepackage{bm}
\usepackage{subfigure}

\newcommand{\mpc}{\, {\rm Mpc}}

\newcommand{\hmpc}{\, h^{-1} \mpc}
\newcommand{\ihmpc}{\, h\, {\rm Mpc}^{-1}}

\newcommand{\fnl}{{f_{\rm NL}}}
\newcommand{\tfnl}{{\tilde{f}_{\rm NL}}}

\newcommand{\Ptt}{{P_{\theta\theta}}}

\begin{document}

\title{How to measure redshift-space distortions without sample variance}

\author{Patrick McDonald}
\email{pmcdonal@cita.utoronto.ca}
\affiliation{Canadian Institute for Theoretical Astrophysics, University of
Toronto, Toronto, ON M5S 3H8, Canada}
\author{Uro{\v s} Seljak}
\affiliation{Physics and Astronomy Department and Lawrence Berkeley National Laboratory,
University of California, Berkeley, California 94720, USA}
\affiliation{Institute for Theoretical Physics, University of Zurich, Switzerland}

\date{\today}

\begin{abstract}

We show how to use multiple tracers of large-scale density with different 
biases to measure the redshift-space distortion parameter 
$\beta\equiv b^{-1}f\equiv b^{-1}d\ln D/d\ln a$ (where $D$ is the growth rate 
and $a$ the expansion factor), to a much better precision than one could 
achieve with a single tracer, to an arbitrary precision in the low noise limit.
In combination with the power spectrum of the tracers this allows a much more 
precise measurement of the bias-free velocity divergence power spectrum, 
$f^2 P_m$ -- in fact, in the low noise limit $f^2 P_m$ can be measured as well 
as would be possible if velocity divergence was observed directly, with rms 
improvement factor 
$\sim\left[5.2\left(\beta^2+2\beta+2\right)/\beta^2\right]^{1/2}$ (e.g., 
$\simeq 10$ times better than a single tracer for $\beta=0.4$). This would 
allow a high precision determination of $f D$ as a function of redshift with an 
error as low as 0.1\%. We find up to two orders of magnitude improvement in 
Figure of Merit for the Dark Energy equation of state relative to Stage II, a 
factor of several better than other proposed Stage IV Dark Energy surveys. The 
ratio $b_2/b_1$ will be determined with an even greater precision than $\beta$,
producing, when measured as a function of scale, an exquisitely sensitive probe
of the onset of non-linear bias. We also extend in more detail previous work on
the use of the same technique to measure non-Gaussianity. Currently planned 
redshift surveys are typically designed with signal to noise of unity on scales
of interest, and are not optimized for this technique. Our results suggest that
this strategy may need to be revisited as there are large gains to be achieved 
from surveys with higher number densities of galaxies. 

\end{abstract}

\pacs{98.65.Dx, 95.35.+d, 98.80.Es, 98.80.-k}

\maketitle

\section{Introduction}

Growth of structure in the universe has long been recognized as one of the 
most powerful ways to learn about the nature of dark energy and other 
properties of our universe. Currently the most promising method is weak 
lensing tomography, which traces the dark matter directly and can measure the 
growth of structure by splitting the source galaxies by their (photometric)
redshift. For example, the Dark Energy Task Force concludes that of all 
proposed next 
generation experiments weak lensing holds the best promise to succeed in 
improving our knowledge of dark energy \cite{2006astro.ph..9591A}. However, 
weak lensing is 
not necessarily the ideal method: it measures the structure projected over a 
broad 
window in redshift and its ability to probe rapid changes in growth rate is 
limited. 
Moreover, one is measuring a 2-dimensional projection only as opposed to the 
full 
3-dimensional information, drastically reducing the amount of available 
information. 
In addition, for all of the methods proposed so far the requirements on 
systematic error 
exceed what is achievable today, so it is worth pursuing multiple methods until
we have a better control of systematics. 

Galaxy clustering has been the favorite method of measuring large scale 
structure in 
the universe and it is likely this will continue also in the future. 
The main reason is that galaxies are easily observed and that by measuring 
their redshift 
one can reconstruct the 3-dimensional clustering information, in contrast to 
weak lensing 
or cosmic microwave background anisotropies which only measure a 2-dimensional 
projection. 
The relation between galaxy and dark matter clustering 
is however not straight-forward. In the simplest model of linear bias galaxies 
trace 
dark matter up to an overall constant called linear bias. The bias cannot be 
predicted 
from the theory and as a result galaxy clustering alone 
cannot measure the growth of structure with redshift. 
If one could determine bias with sufficiently small error then galaxy 
clustering 
would become the leading method due to its higher information content. 

There are several methods proposed to determine the bias. 
One is to use weak lensing, specifically 
galaxy-galaxy lensing which measures the cross-correlation between galaxies and
dark matter. 
This is proportional to bias $b$ and in combination with the galaxy 
auto-correlation 
function which scales as $b^2$ one can eliminate the dependence on bias to 
measure
$\Omega_{m 0}^2P_m(k)$, where $\Omega_{m0}$ is the matter density parameter 
today and 
$P_m(k)$ is the matter power spectrum at a given redshift 
\cite{2002ApJ...577..604H,2004AJ....127.2544S}. 
Alternatively, one can also measure the halo mass distribution with weak 
lensing 
which, in connection with the 
theoretical bias predictions, can determine the bias and thus $P_m(k)$ 
\cite{2005PhRvD..71d3511S}. 
A second method to determine the bias and thus $P_m(k)$ is to measure the 
three-point function \cite{2002MNRAS.335..432V}. 
A third method, and the one we focus here, is to measure the redshift space 
distortion 
parameter 
$\beta \equiv b^{-1} f\equiv b^{-1} d\ln D / d\ln a $ (where 
$b$ is the bias of the galaxies, $D$ is the linear theory growth factor, and
$a$ is the expansion factor) 
\cite{2003MNRAS.346...78H,2004MNRAS.353.1201P,2006PhRvD..74l3507T,
2008Natur.451..541G}. In combination with the galaxy power spectrum 
this gives $f^2 P_m$. Note that these methods give somewhat different dynamical 
measurements once the bias is eliminated, so to some extent they are 
complementary to each other. However, 
none of these methods is presently competitive in terms of 
derived cosmological constraints, as they all have rather large statistical 
errors
from the current data, although this may change in the future as data improve 
and 
new analysis methods are developed
\cite{2006ApJ...652...26Y,2006PhRvD..74b3522S,2008arXiv0807.0810S,
2008arXiv0808.0003P}. 

In this paper we focus on redshift space distortion parameter $\beta$ as a way 
to 
determine the bias. Clustering of galaxies along the line of sight is enhanced 
relative to the transverse direction due to peculiar motions and this allows 
one to 
determine $\beta$. Current methods require one to compare the clustering 
strength as a
function of angle relative to the line of sight, but this method is only 
applicable on large 
scales where linear theory holds. As a result, the sampling variance limits its
statistical precision. 
Recently, \cite{2008arXiv0807.1770S} proposed a new method for 
measuring the primordial non-Gaussianity parameter $\fnl$, by comparing two
sets of galaxies with a different bias, and a different sensitivity to $\fnl$
\cite{2008PhRvD..77l3514D},
which allows one to eliminate the sample variance. 
Optimization of this technique was investigated by \cite{2008arXiv0808.0044S}.
Here we apply this technique to the measurement of the redshift-space 
distortion parameter, which in turn improves
the measurement of the velocity divergence power spectrum 
$\Ptt\equiv f^2 P_m(k)$.
As we show here, this approach can in principle measure $\beta$ perfectly and 
$\Ptt$ as well as if we observed velocity divergence directly. 
If this promise were realized from the data it would allow for a much higher 
statistical 
power than weak lensing or other 2-dimensional projections. 
We also show that our method, and large-scale structure surveys in general, 
become even more powerful 
when additional cosmology dependence, such as the Alcock-Paczy{\'n}ski 
effect \cite{1979Natur.281..358A}, is included.

We begin by presenting the basic method, followed by
an analysis of the expected improvement as a function of survey parameters. 
This is followed by 
predictions for some of the existing and 
future surveys in terms of expected improvement of dark energy 
parameters.  On small scales 
the deterministic linear bias model
eventually becomes inaccurate and for this reason we often quote results as
a function of the maximum usable wavenumber, $k_{\rm max}$. 
While we have some idea what its value should be, we leave a more detailed 
analysis with numerical simulations for the future. 
One should note,
however, that the multiple-tracer approach is likely to actually help in 
disentangling non-linear bias effects on quasi-linear scales, making those
scales potentially much more useful than they would otherwise be.
We conclude with a summary and a discussion of future directions. 

\section{Method and Results}

The density perturbation, $\delta_{g i}$, for a type of galaxy $i$, in the 
linear regime, in redshift space, is \cite{1987MNRAS.227....1K} 
\begin{equation}
\delta_{g i}=\left(b_i+f\mu^2\right) \delta +\epsilon_i
\label{eqkaiser}
\end{equation}
where $b_i$ is the galaxy bias, $\mu=k_\parallel/k$, $\delta$ is the mass 
density
perturbation, and $\epsilon_i$ is a white noise variable which can 
represent either 
the standard shot-noise or other stochasticity.  
All equations are understood to apply to the real or imaginary part of a 
single Fourier mode unless otherwise indicated.
In this paper we will consider two types of galaxies, type 1 with bias $b$, 
and type 2 with bias $\alpha b$.  
The perturbation equations can then be written
\begin{equation}
\delta_{g 1}=f \left(\beta^{-1}+ \mu^2\right) \delta +\epsilon_1~,
\label{eq1}
\end{equation}
and
\begin{equation}
\delta_{g 2}=f \left(\alpha\beta^{-1}+ \mu^2\right) \delta +\epsilon_2~.
\label{eq2}
\end{equation}
We are denoting $\beta=f/b$ (the equivalent distortion parameter for the 2nd 
type of galaxy is $\beta/\alpha$). 

The traditional method to determine $\beta$ is to look at the angular 
dependence
of the two-point correlation function or its Fourier transform, the power 
spectrum. The correlations 
will be enhanced along the line of sight (where $\mu=1$) 
relative to the direction perpendicular to it (where $\mu=0$), 
but to observe this enhancement one must average over many independent modes to 
beat down the sampling (or cosmic) variance. This is because each mode is a 
random realization of a Gaussian field and there will be fluctuations in the 
measured power even in the absence of noise. By combining a measurement of 
$\beta$
with that of the galaxy power spectrum one can determine $f^2 P_m$, which no 
longer 
depends on the unknown bias of the galaxies. This method has been applied to 
the 
data, most recently in \cite{2003MNRAS.346...78H,2004MNRAS.353.1201P,2006PhRvD..74l3507T,2008Natur.451..541G},
and is limited by the accuracy with which we can 
determine $\beta$. For example, for the analysis in \cite{2006PhRvD..74l3507T},
which currently has the highest signal to noise 
measurement of $\beta$, the error on the overall amplitude of 
galaxy power spectrum $P_{gg}$ is about 1\% adding up all the modes up to $k=0.1\ihmpc$, 
while the error on $\beta$ is about 12\%, so the 
error on reconstructed $f^2 P_m=\beta^2 P_{gg}$ is entirely dominated by the 
error on $\beta$. Predictions 
for the future surveys with this and related methods can be found in \cite{2008arXiv0807.0810S,2008arXiv0808.0003P}. 

To understand the main point of the new method we are proposing here 
let us consider the situation without the noise, which would apply if, for example, 
we have a very high density of the two tracers sampling the field, and no 
stochasticity. In that case 
we can divide equation \ref{eq2} by equation \ref{eq1} above to obtain
\begin{equation}
{\delta_{g 2} \over \delta_{g 1}}={ \alpha\beta^{-1}+ \mu^2 \over \beta^{-1}+ \mu^2}.
\end{equation}
This expression has a specific angular dependence, allowing one to extract  $\alpha$ and 
$\beta$ separately. Note that there is no dependence on the 
density field $\delta$. The random nature of the density field is therefore not 
affecting this method and we can determine $\beta$ exactly in the absence of noise. 
More generally, the precision with which we can determine $\beta$ is controlled by the (shot) 
noise, i.e. density of tracers, rather than the sampling variance. 
In order to address the gains in a realistic case 
we must perform the full analysis, which we turn to next. 

Generally, the noise variables can be correlated with each other, although 
they would not be for standard shot-noise. For example,
\cite{2006PhRvD..74j3512M} showed that one generically expects non-linear 
structure formation to generate a full covariance matrix for the noise at some 
level, so this will probably be the
ultimate limit for this kind of measurement.
The covariance matrix of the perturbations is 
\begin{equation}
C \equiv \left[ \begin{array}{cc}
\left<\delta^2_{g 1}\right> & \left<\delta_{g 1} \delta_{g 2}\right>     \\
 \left<\delta_{g 2} \delta_{g 1}\right> & \left<\delta^2_{g 2}\right>
\end{array} \right] = \frac{\Ptt}{2} \left[ \begin{array}{cc}
\left(\beta^{-1}+\mu^2\right)^2 & \left(\beta^{-1}+\mu^2\right)
  \left(\alpha\beta^{-1}+\mu^2\right)   \\
\left(\beta^{-1}+\mu^2\right)   \left(\alpha\beta^{-1}+\mu^2\right)  &
\left(\alpha\beta^{-1}+\mu^2\right)^2 
\end{array} \right] + \frac{N}{2}
\label{eqcovarmat}
\end{equation}
where $\Ptt\equiv 2 f^2\left<\delta^2\right>$ and 
$N_{ij}\equiv 2 \left<\epsilon_i \epsilon_j\right>$ (note that $\Ptt$ and $N$
are the
usual power spectrum and noise -- the factor of 2 comes from the fact that 
$\delta$ and $\epsilon$ are only the real or imaginary part of a Fourier mode).   
If we assume that the noise matrix is known then the covariance matrix for two types of galaxies is 
a function of
three parameters:  the velocity divergence power spectrum amplitude, $\Ptt$, 
the 
redshift-space distortion parameter, $\beta$, and the ratio of biases 
$\alpha$.  
We work with $\alpha$ instead of a second $\beta$ parameter because 
the ratio of biases will generally be determined substantially 
more precisely than $\beta$, which means that the measurement of the 
second $\beta$ parameter would be almost perfectly correlated with the first.
The Fisher matrix for the measurement of these parameters is
\begin{equation}
F_{\lambda\lambda^\prime}=\frac{1}{2}{\rm Tr}\left[C_{,\lambda} C^{-1} 
C_{,\lambda^\prime} C^{-1}\right]
\label{eqfishmat}
\end{equation}
where $C_{,\lambda}\equiv dC/d\lambda$ and $\lambda$ are the parameters. 

For any single mode, the inverse of the Fisher matrix is singular, i.e., we
can only constrain two parameters, not three.  However, adding Fisher matrices
for modes with different $\mu$ breaks this degeneracy.
Generally, the total Fisher matrix will be an integral over modes with all 
angles. 
As a simple example, we assume first a pair of modes with $\mu=0$ and $\mu=1$ 
and 
compute the error on the parameters in the small noise limit:
\begin{equation}
\frac{\sigma^2_\alpha}{\alpha^2} = X_{11}- 2 X_{12}+X_{22}~,
\end{equation}
where $X_{ij}=N_{ij}/b_i b_j P_m$, 
\begin{equation}
\frac{\sigma^2_{\beta}}{\beta^2}=
\frac{\left[\alpha^2 \left(1+\beta\right)^2+
\left(\alpha+\beta\right)^2\right]X_{11}
-2 \left[\alpha^2\left(1+\beta\right)^2+
\alpha\left(1+\beta\right)\left(\alpha+\beta\right)\right] X_{12}
+2 \alpha^2\left(1+\beta\right)^2 X_{22}}{\beta^2\left(\alpha-1\right)^2}~,
\end{equation}
and
\begin{equation}
\frac{\sigma^2_\Ptt}{\Ptt^2} = 1~.
\end{equation}
The key points are that the only lower limit on the errors on $\alpha$ and 
$\beta$ from this
single pair of modes is set by the achievable noise-to-signal ratios on the 
tracers, $X_{ij}$, and
the error on $\Ptt$ is the error one would obtain for a simple power spectrum 
measurement from two modes, with no degradation due to degeneracy with the bias
parameters (this is of course only true to leading order in the small noise 
limit).  

In contrast, for a single type of
galaxy, for the same pair of radial and transverse modes, in the low noise
limit we find: 
\begin{equation}
\frac{\sigma^2_\beta}{\beta^2} =\frac{\left(1+\beta\right)^2}{\beta^2}
\end{equation}
and
\begin{equation}
\frac{\sigma^2_\Ptt}{\Ptt^2} = 
\frac{2 \left(\beta^2+2 \beta+2\right)}{\beta^2}
\end{equation}
i.e., here the signal-to-noise per mode on $\beta$ saturates at 
greater than unity for low noise, and the overall power measurement is
substantially degraded because we cannot determine $\beta$ very accurately and this 
then limits the precision with which the velocity divergence power spectrum can 
be determined as well.
If we assume the first tracer is much better sampled than the second, we get
the ratio of errors
\begin{equation}
\frac{\sigma^2_{\beta}\left(2~{\rm tracers}\right)}
{\sigma^2_{\beta}\left(1~{\rm tracer}\right)} \simeq
\frac{2 \alpha^2 X_{22}}{\left(\alpha-1\right)^2}=
\frac{2 N_{22}}{\left(b_2-b_1\right)^2 P_m}
\label{eqrelativebetaerror}
\end{equation}
and
\begin{equation}
\frac{\sigma^2_\Ptt\left(2~{\rm tracers}\right)}
{\sigma^2_\Ptt\left(1~{\rm tracer}\right)} \simeq
\frac{\beta^2}{2 \left(\beta^2+2 \beta+2\right)}~.
\label{eqrelativePerror}
\end{equation}
We see that the improvement in the measurement of $\beta$ can be arbitrarily 
large in the limit $N_{22} \rightarrow 0$, 
while the improvement in $\Ptt$ is limited by sample variance but is 
generally quite large, e.g., a factor $10^{1/2}$ (rms) for $\beta=1$, and 
$50^{1/2}$ for $\beta=1/3$.
All of these formulae are only good as long as the noise is small relative to
the power spectrum of the difference between the two tracer fields 
(obviously the two tracer case should never be worse than the one tracer 
case). 

A yet more realistic analysis must integrate over all modes rather than look 
just at the two specific modes as done above. In doing so we find that 
the two-mode approximation above substantially 
underestimates the improvement in the $\Ptt$ measurement.  If we integrate the 
Fisher matrix over the full range of $\mu$, the improvement is a factor of
approximately $2.6^{1/2}$ larger than above, e.g., a factor of $25^{1/2}$
for $\beta=1$, or $138^{1/2}$ for $\beta=1/3$.  
The result of the $\mu$ integration in the idealized case of two perfect tracers remains
simply:
\begin{equation}
\frac{\sigma^2_\Ptt}{\Ptt^2} = 
\frac{2}{N},
\end{equation}
where $N$ is the effective number of discrete modes sampled (counting real and
imaginary parts separately). Thus, the
combination of two perfect tracers produces the measurement of $f^2 P_m$ that 
one would obtain if one could observe the velocity divergence field directly
and perfectly.

\subsection{General usefulness of redshift-space distortion measurements}

Of course, we will never have perfect sampling, so it is necessary to 
numerically evaluate
the exact Fisher matrix with finite noise to determine the improvement for
real surveys.  
We assume the survey volume, $V$, is roughly spherical, so the
Fisher matrix for a survey will be
\begin{equation}
F_V\simeq \frac{V}{4 \pi^2} 
\int_{k_{\rm min}}^{k_{\rm max}} k^2 dk \int_{-1}^1 d\mu ~ F\left(k,\mu\right) 
\end{equation}
where $F\left(k,\mu\right)$ is computed for a single mode using Eqs. 
(\ref{eqcovarmat}) and (\ref{eqfishmat}) (remembering that both real and
imaginary parts must be added).
We assume the minimum usable $k$ is $k_{\rm min}\simeq 2 \pi / V^{1/3}$.

\begin{figure}
\subfigure{\includegraphics[width=0.49\textwidth]{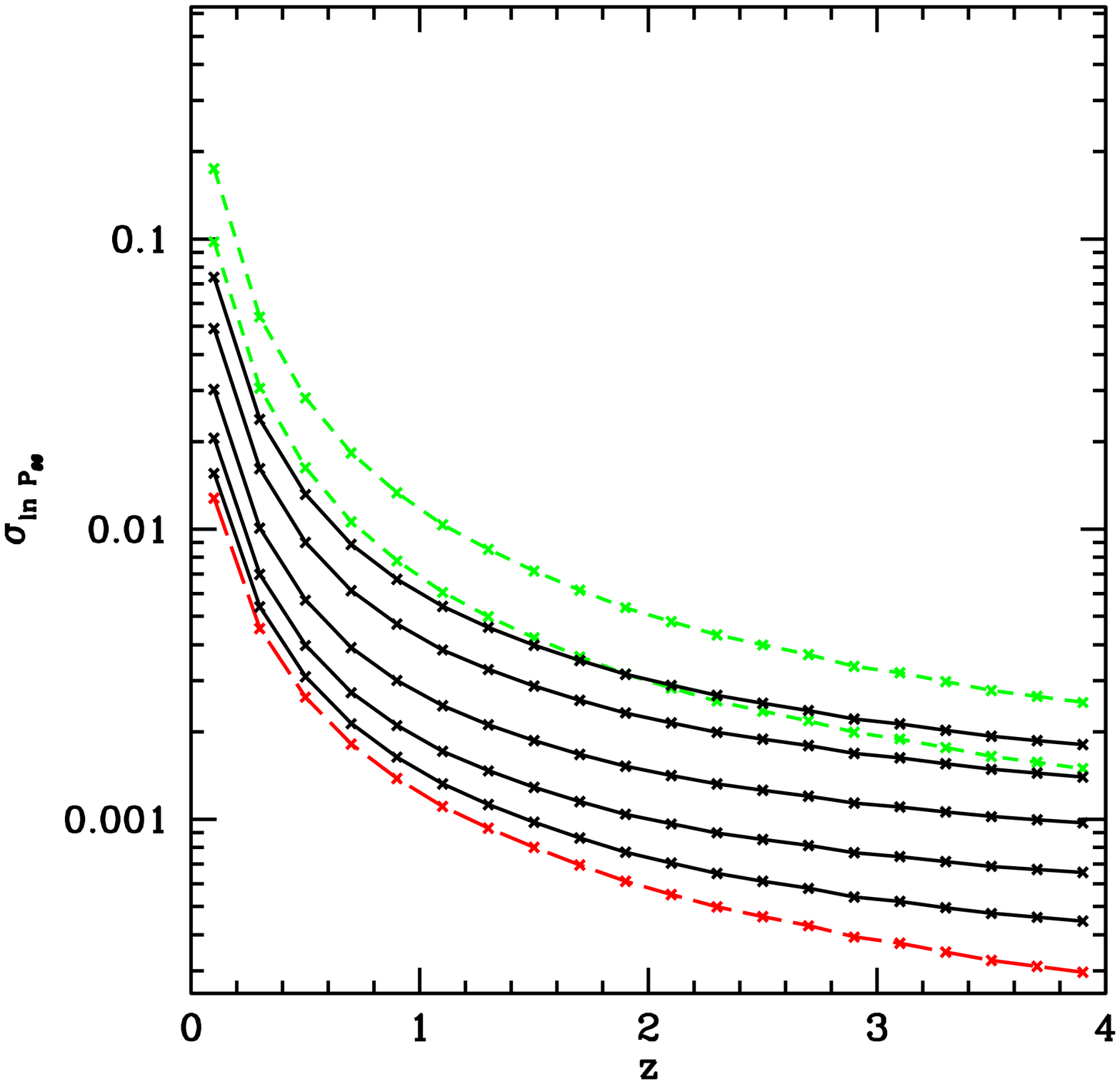}}
\subfigure{\includegraphics[width=0.49\textwidth]{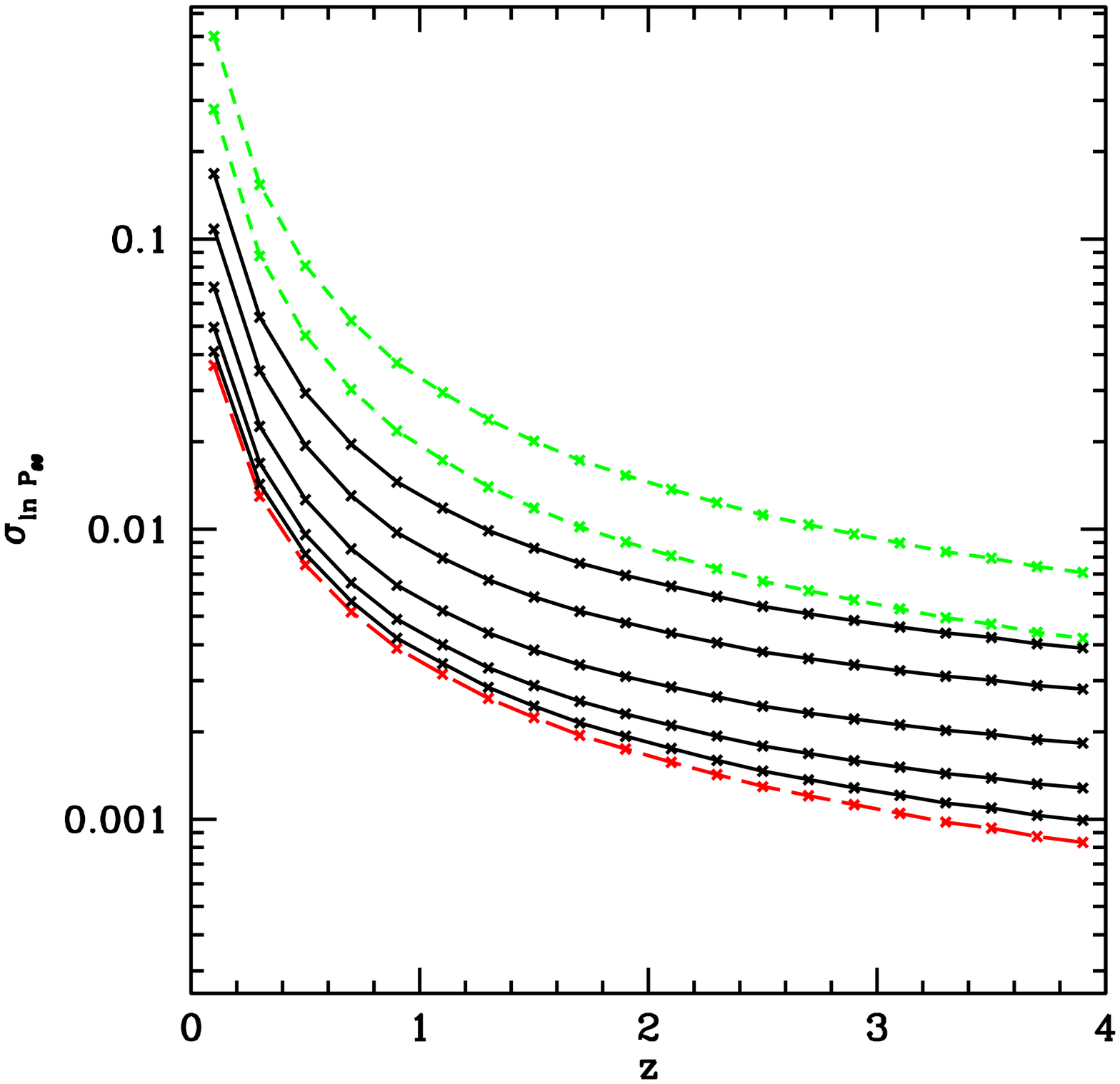}}
\caption{
Projected fractional error on the normalization of 
$\Ptt\equiv f^2 P_m$, for 30000 square degrees, in redshift bins
with width $dz=0.2$. 
The upper and lower green (short-dashed) lines show the constraints from 
single tracers with $b=2$ and $b=1$, respectively. 
Black (solid) lines show the two tracers together, each with, from top
to bottom, 
S/N=1, 3, 10, 30, 100  at $k=0.4\ihmpc$. 
The red (long-dashed) line shows the case where both
tracers are perfectly sampled.
For the left panel we assume $k_{\rm max}(z)=
0.1~\left[D\left(z\right)/D\left(0\right)\right]^{-1}\ihmpc$, while the 
right assumes $k_{\rm max}(z)=
0.05~\left[D\left(z\right)/D\left(0\right)\right]^{-1}\ihmpc$.
}
\label{figf2Pvsz}
\end{figure}

In Figure \ref{figf2Pvsz} we show the errors on the amplitude of the
velocity divergence power spectrum $\Ptt$ that one
can obtain from a generic 3/4-sky (30000 square degree) survey, as a function
of redshift. 
This may seem unreasonably optimistic/futuristic, but it may be possible 
surprisingly quickly if the 21 cm emission intensity mapping method is 
successful
\cite{2008PhRvL.100i1303C}, as discussed further below.
We assume either 
$k_{\rm max}(z)=
0.1~\left[D\left(z\right)/D\left(0\right)\right]^{-1}\ihmpc$, or
$k_{\rm max}(z)=
0.05~\left[D\left(z\right)/D\left(0\right)\right]^{-1}\ihmpc$,
with the redshift dependence motivated
by \cite{2006PhRvD..73f3520C,2007ApJ...665...14S}. 
We give results for different values of 
$X_{ij}$ at $k=0.4\ihmpc$ (for $i=j$, i.e., the off-diagonal element is zero). 
Surveys with S/N$\sim 1$ at $k=0.4 \ihmpc$ recover $\sim 90$\% of the 
total possible information from BAO measurements, i.e., the S/N ratios we give
should be interpreted as being relative to a nearly ideal BAO survey.
In these calculations we are assuming the shape of the power spectrum 
is known so that we can compress the information from all the modes into a 
single 
number, that of the power spectrum amplitude. This is discussed further below. 

We see from Fig. \ref{figf2Pvsz} that one obtains large improvements using the 
multi-tracer method, 
even for S/N as small as 3. For example,
at $z\sim 1$, with $k_{\rm max}(z)=
0.1~\left[D\left(z\right)/D\left(0\right)\right]^{-1}\ihmpc$,
the gain is equivalent to increasing the survey volume by a factor
of 3 over a single, perfectly sampled, unbiased tracer. 
For the more conservative $k_{\rm max}(z)$,
the results are substantially degraded,
but the multi-tracer method actually 
becomes more valuable, at fixed absolute S/N level, because the S/N level in
the usable range of $k$ is higher. 

It is interesting to note that, in the single-tracer case, low bias is 
substantially better than high bias. This is easy to understand from Eq. 
\ref{eqkaiser} -- lower bias simply means that
power due to redshift-space distortions makes a larger fractional contribution
to the total. The practical consequence of this observation is that,
once BAO surveys using high bias objects have been completed, an opportunity
exists for improvement by observing lower bias objects at similarly low S/N, 
i.e., 
even if the S/N is not large enough to decisively 
exploit the multi-tracer method in this
paper. 

\begin{figure}
\resizebox{\textwidth}{!}{\includegraphics{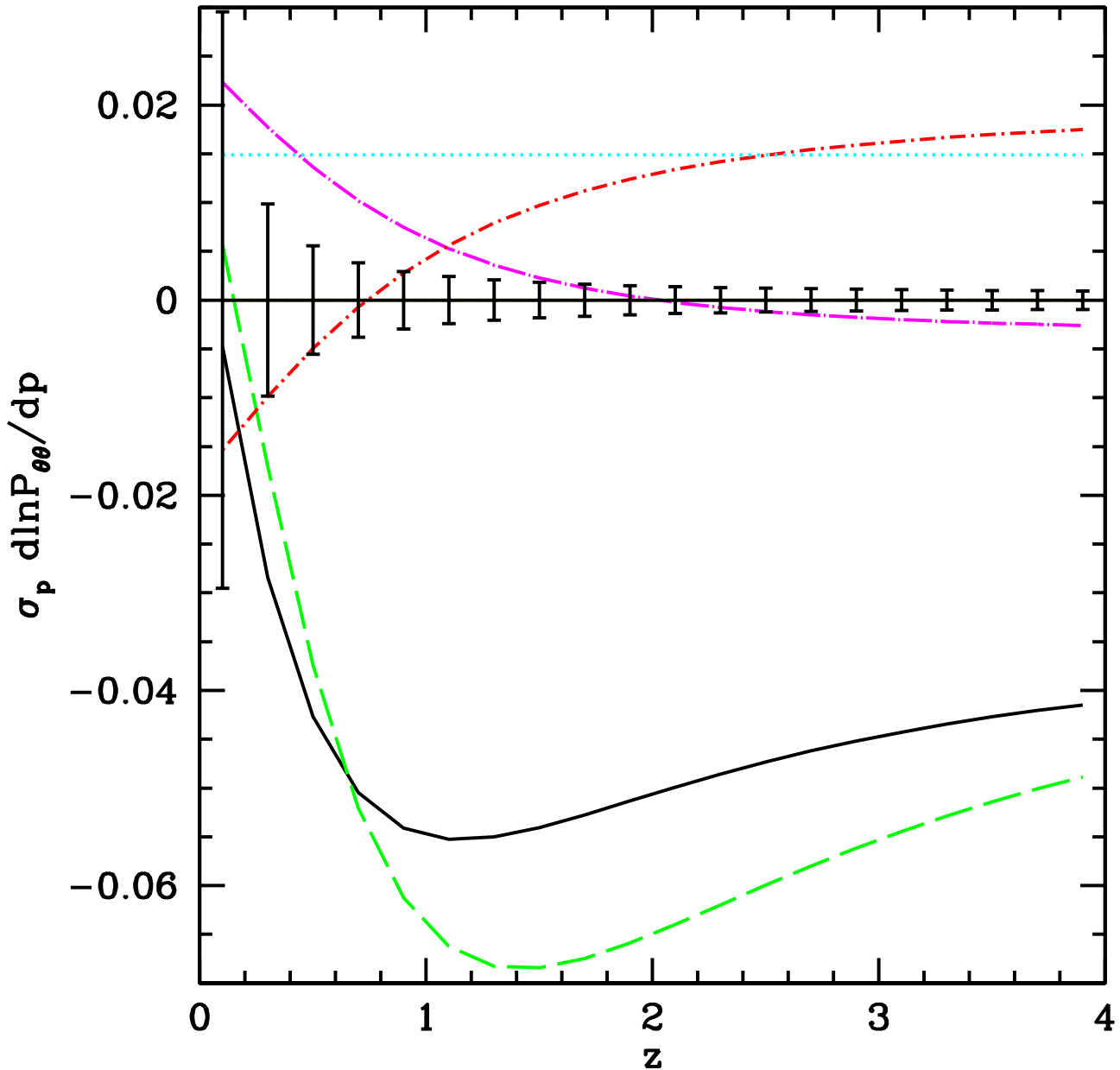}}
\caption{
Dependence of $\Ptt\left(k=0.1\ihmpc,z\right)$ on cosmological parameters.
The derivatives are at fixed values of the other parameters, including 
not-shown parameters $\omega_b$, $\theta_s$, and $n_s$. The derivatives are
normalized by typical projected errors on the parameters, for the scenario 
described in Fig. \ref{figallparamsvsz}, with $z_{\rm max}=1$.
Lines correspond to:
black (solid): $w_0$,
green (long-dashed): $w^\prime$,
red (dot-short-dashed): $\Omega_k$,
magenta (dot-long-dashed): $\omega_m$,
cyan (dotted): $\log A$.
Error bars show the projected errors on $\Ptt$ for
$k_{\rm max}(z)=
0.1~\left[D\left(z\right)/D\left(0\right)\right]^{-1}\ihmpc$, and $S/N=10$. 
}
\label{figdPdp}
\end{figure}

\begin{figure}
\resizebox{\textwidth}{!}{\includegraphics{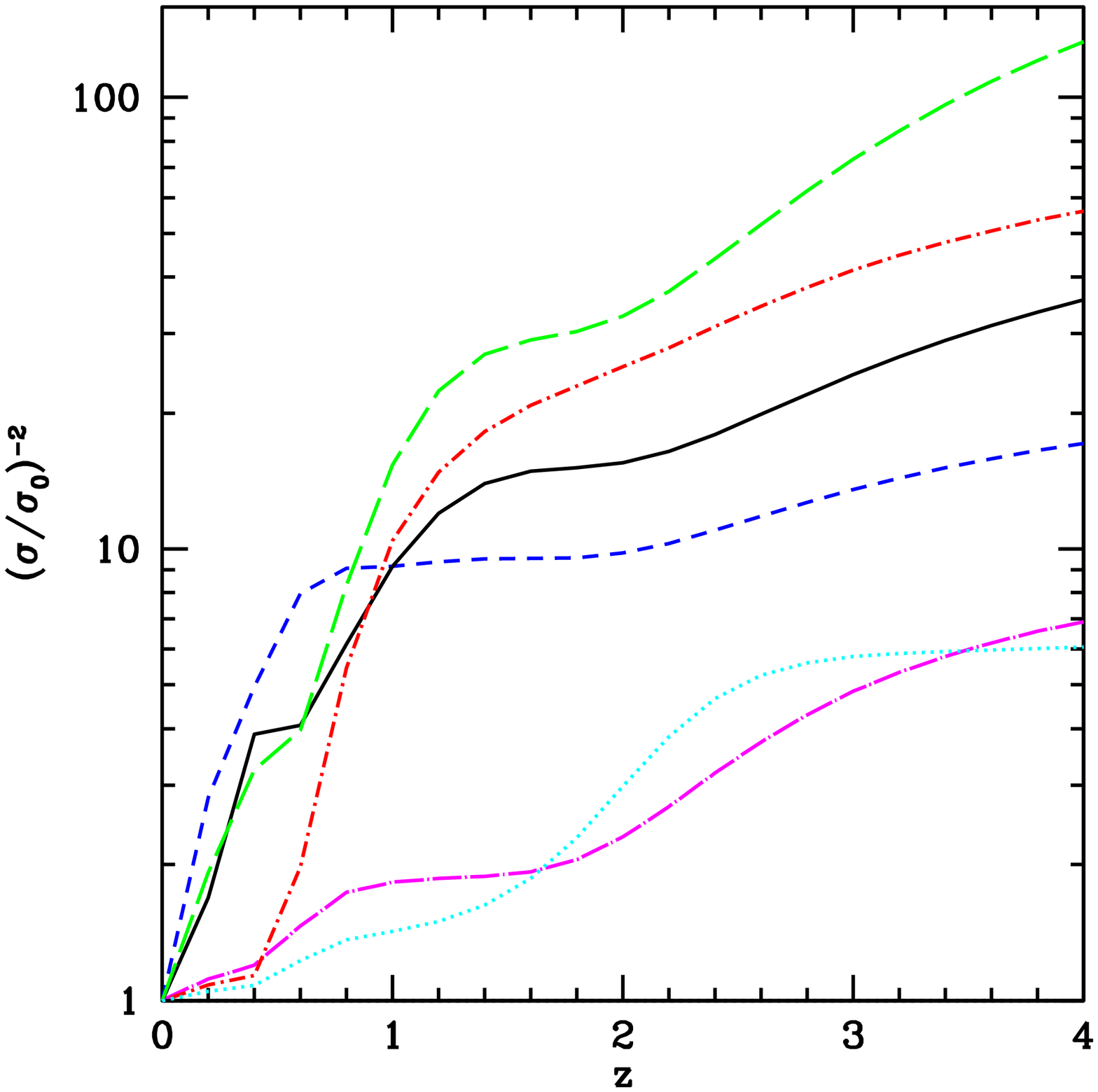}}
\caption{
Improvements in constraints on various parameters, vs. $z_{\rm max}$, 
for the case including BAO, 
 $k_{\rm max}(z)=
0.1~\left[D\left(z\right)/D\left(0\right)\right]^{-1}\ihmpc$, 
and $S/N=10$ for the $\Ptt$ 
measurement.  Lines are:
black (solid): $w_0$,
green (long-dashed): $w^\prime$,
red (dot-short-dashed): $\Omega_k$,
blue (dashed): $\Omega_\Lambda$,
magenta (dot-long-dashed): $\omega_m$,
cyan (dotted): $\log A$.
}
\label{figallparamsvsz}
\end{figure}

To put these results into a broader context, Fig. \ref{figdPdp}
shows how velocity divergence power spectrum $\Ptt=f^2 P_m$ depends on the 
underlying cosmological parameters of interest,
as a function of redshift.
The derivatives with respect to parameter $p_i$, $d \ln \Ptt/dp_i$,
are taken at fixed values of the other parameters in the following set:
$\omega_m=\Omega_m h^2$, the matter density; $\omega_b=\Omega_b h^2$, the
baryon density; $\theta_s$, the angle 
subtended by the sound horizon at the CMB decoupling redshift; $w_0$ and
$w^\prime$, defined by $w(z)=w_0 + w^\prime \left(1-a \right)=p(z)/\rho(z)$
where here $p(z)$ and $\rho(z)$ are the dark energy pressure and density;
$\Omega_k$, the effective fraction of the critical density in curvature; 
and $A$ and $n_s$, the amplitude and
power law slope of the primordial perturbation power spectrum. 
Fixing $\omega_m$, $\omega_b$,
$\theta_s$, $A$, and $n_s$ guarantees that changes in $w_0$, $w^\prime$,
and $\Omega_k$ have minimal effect on the CMB. The scale for the parameter
dependences is chosen using the error bars on each parameter that we project 
for
the scenario described for Fig. \ref{figallparamsvsz}, including the Planck
CMB experiment, and $\Ptt$ and BAO constraints up to $z_{\rm max}=1$. 
While it is easy to see that $\Ptt$ measurements have a lot of 
power to
constrain parameters, especially the dark energy equation of state, exactly 
what constraints can be obtained is not obvious to the eye, 
because one needs to account
for degeneracy between parameters.

In the derivatives in Fig. \ref{figdPdp}, 
and the following calculations, we assume that the 
constraints on $\Ptt$ can be taken as a measurement of 
$\Ptt$ at $k=0.1\ihmpc$, for all $z$ and all cosmological models.
This allows us to apply the constraints directly to the product $f(z) D(z)$, 
with the power spectrum entering only through a single overall normalization. 
We do this for simplicity and transparency (i.e., so one can see clearly where
the constraints are coming from), but 
it is imperfect in two ways: first, the 
measurement at different redshifts is really weighted toward different $k$, so
the shape of the power spectrum enters in converting between them; and,
second, we can
not measure the power spectrum in $\hmpc$ units, only radial velocity 
separations and angular separations, which leads to sensitivity to 
the Hubble parameter, $H(z)$, and angular diameter distance, $D_A(z)$ 
(this includes, but is not entirely, the 
Alcock-Paczy{\'n}ski effect \cite{1979Natur.281..358A}). We will revisit these
issues later.

\begin{figure}
\subfigure{\includegraphics[width=0.49\textwidth]{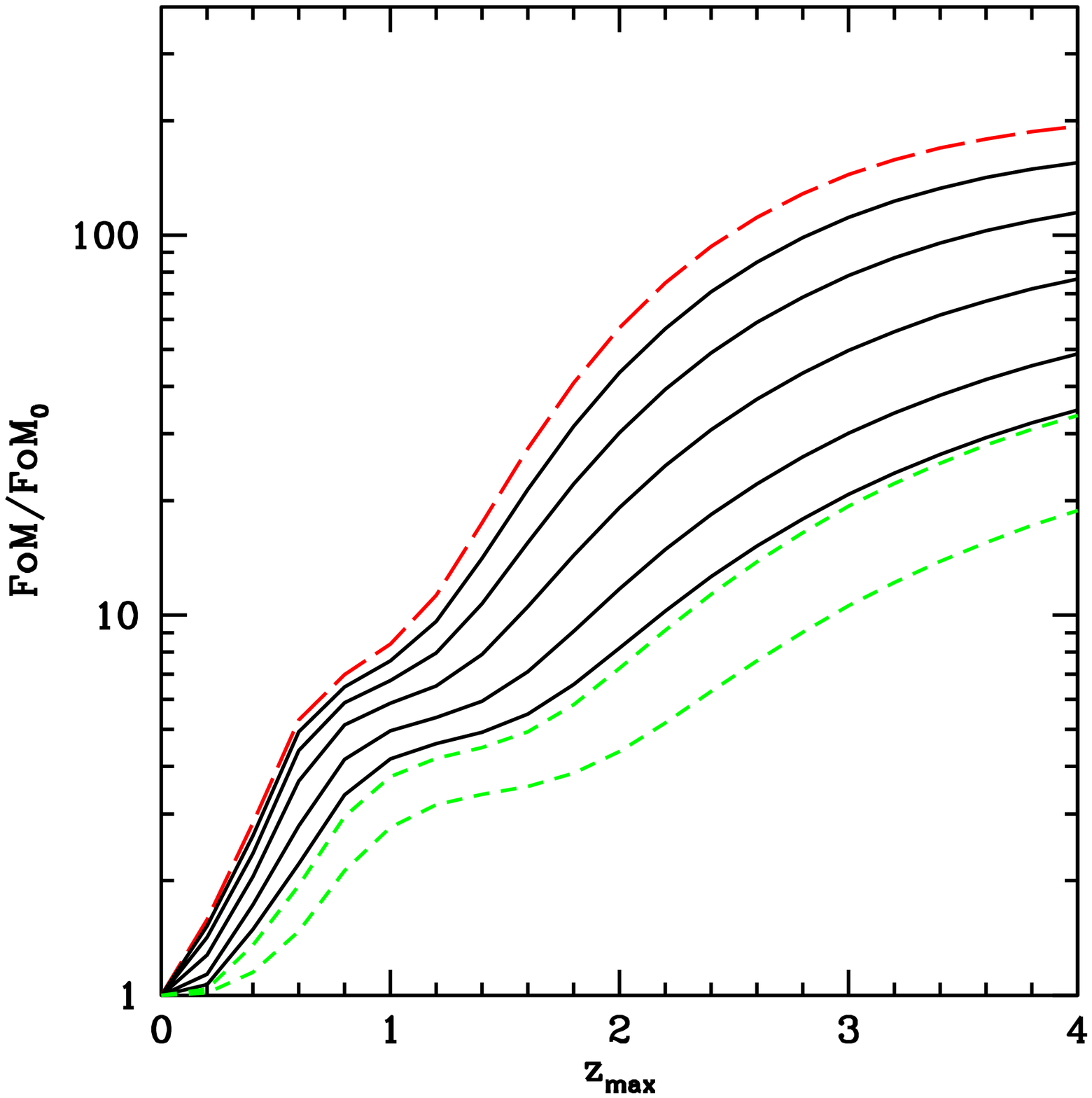}}
\subfigure{\includegraphics[width=0.49\textwidth]{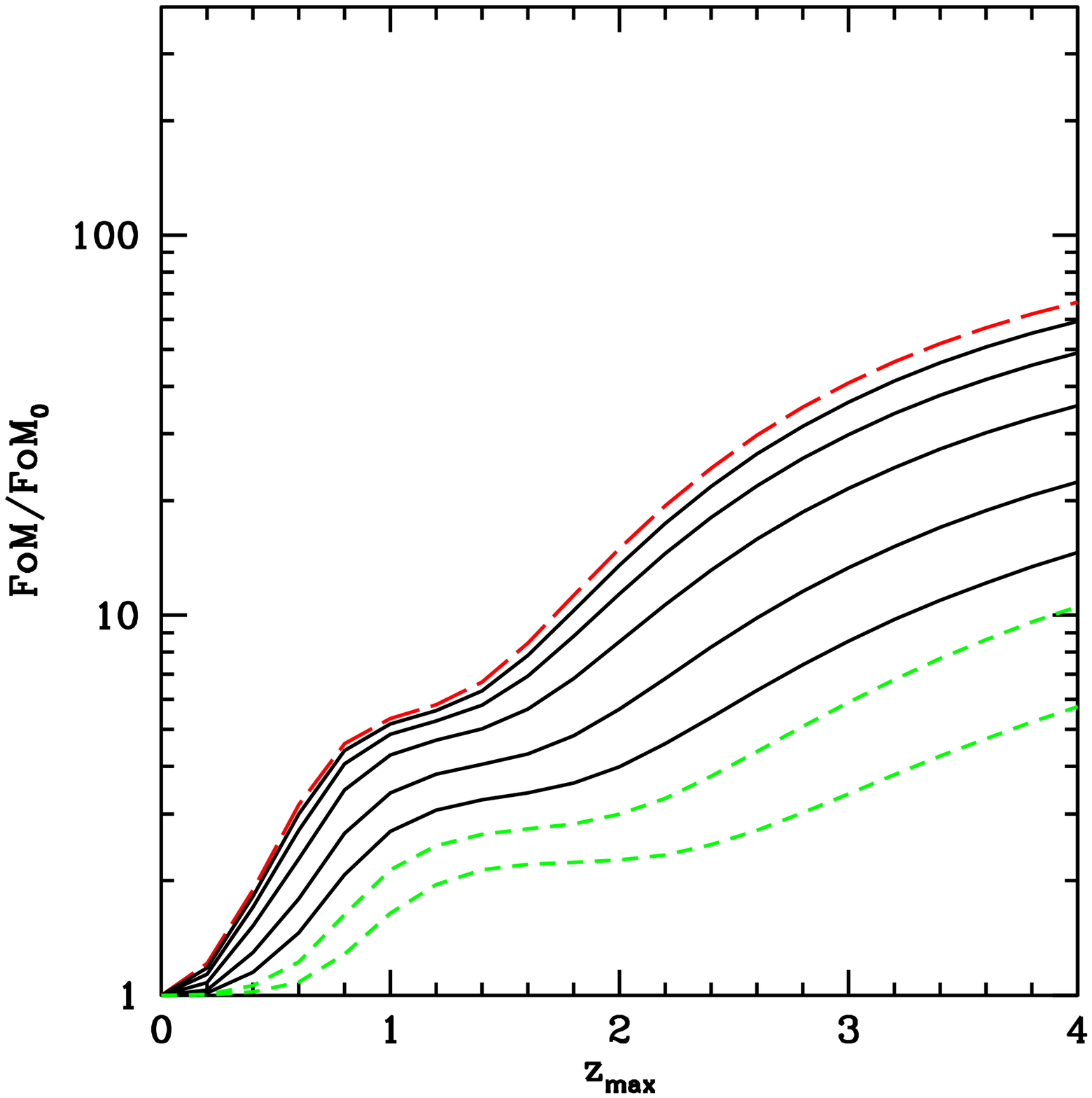}}
\caption{
Improvement in Dark Energy constraints (quantified by the DETF FoM), relative 
to Planck + DETF Stage II, when the 
$\Ptt$ constraints from Fig. \ref{figf2Pvsz} are added.
The lines refer to the same cases as in Fig. \ref{figf2Pvsz}, and again the
left panel is $k_{\rm max}(z)=
0.1~\left[D\left(z\right)/D\left(0\right)\right]^{-1}\ihmpc$ while the 
right is $k_{\rm max}(z)=
0.05~\left[D\left(z\right)/D\left(0\right)\right]^{-1}\ihmpc$.
}
\label{figfom}
\end{figure}

In Fig. \ref{figfom}
we show the contribution these
measurements of $\Ptt$ can make to the study of dark energy,
quantified by the Dark Energy Task Force (DETF) Figure of Merit (FoM), 
proportional to the inverse of the area within the 95\% confidence contours
describing constraints on $w_0$ and $w^\prime$
\cite{2006astro.ph..9591A}.
We see that there is the potential for a huge improvement in dark energy
constraints, by two orders of 
magnitude in the FoM, when $k_{\rm max}(z)=
0.1~\left[D\left(z\right)/D\left(0\right)\right]^{-1}\ihmpc$, well beyond even 
the 
optimistic Stage IV constraints envisioned by the DETF (which gave an 
improvement of a factor $\sim 20$ in the units of this figure).   
The FoM improvements for the $k_{\rm max}(z)=
0.05~\left[D\left(z\right)/D\left(0\right)\right]^{-1}\ihmpc$ 
constraints are substantially weaker of 
course, but still potentially spectacular.

\begin{figure}
\subfigure{\includegraphics[width=0.49\textwidth]{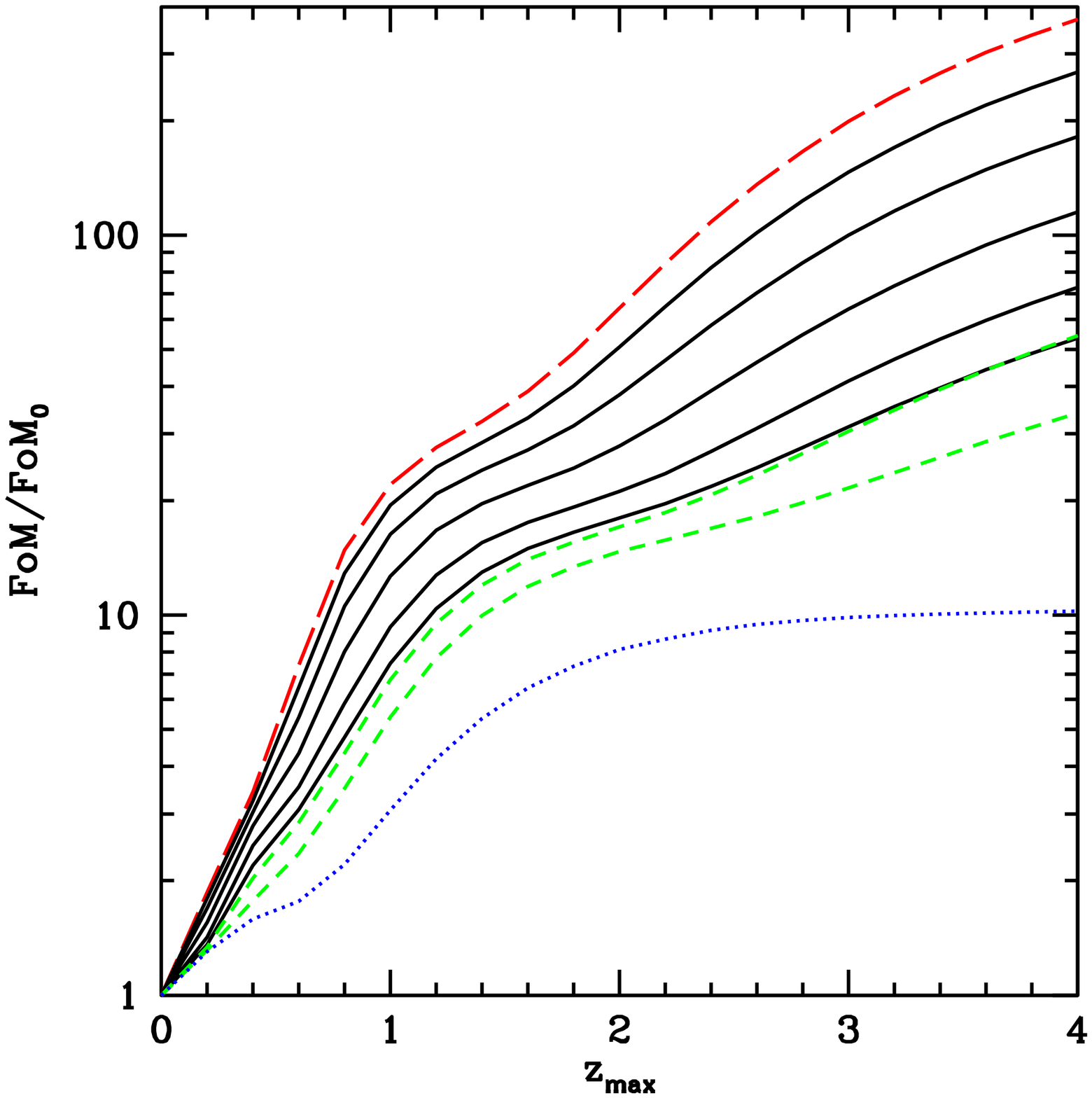}}
\subfigure{\includegraphics[width=0.49\textwidth]{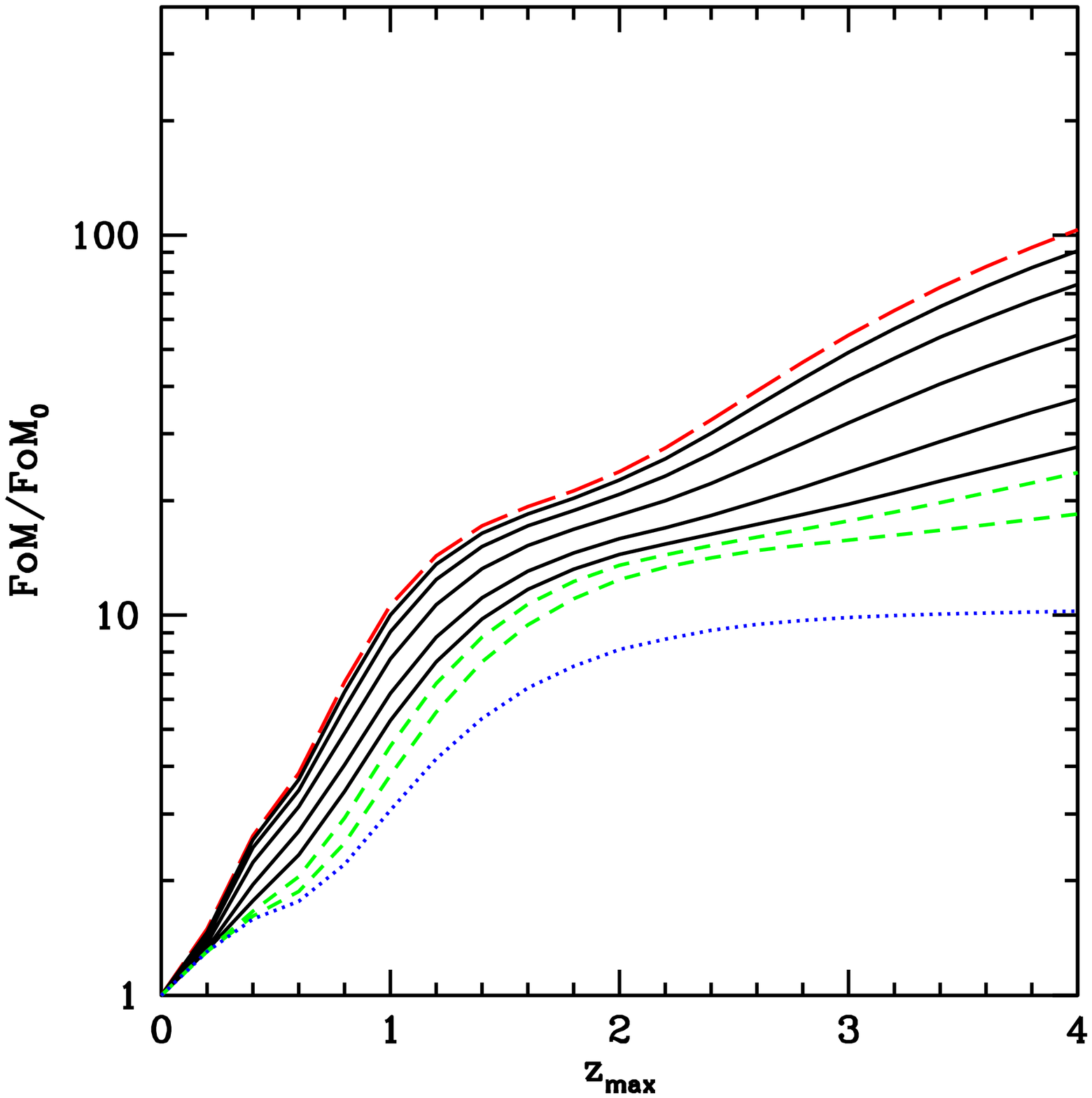}}
\caption{
Improvement in Dark Energy constraints (quantified by the DETF FoM), relative 
to Planck + DETF Stage II, when the 
$\Ptt$ constraints from Fig. \ref{figf2Pvsz} are added, and
the BAO distance measurements from the same survey is added.
The lines refer to the same cases as in Fig. \ref{figf2Pvsz}, except
the blue (dotted) line is the case with BAO alone. The left and right panels
again show the stronger and weaker values of $k_{\rm max}$, respectively.
}
\label{figfomwbao}
\end{figure}

Finally, we add the BAO distance measurements that one would obtain from the
same surveys, with the resulting FoM shown in Fig. \ref{figfomwbao}.
We see that even a minimal $\Ptt$ measurement adds substantially
to the BAO-only measurement.
To give more meaning to these FoM improvement numbers: in the best case
shown here, the constraint on $w(z_p)$ at the pivot point $z_p \simeq 0.8$ 
(where the errors on $w(z_p)$ and $w^\prime$ are uncorrelated) is  
$\pm 0.0023$, with $w^\prime$ is measured to $\pm 0.031$. In the less
optimistic case of $S/N=10$ out to $z_{\rm max}=2$, with 
 $k_{\rm max}(z)=
0.1~\left[D\left(z\right)/D\left(0\right)\right]^{-1}\ihmpc$, 
we find $w(z_p=0.45)$ is constrained
to $\pm 0.0074$, and $w^\prime$ to $\pm 0.091$.

Finally, in Fig. \ref{figallparamsvsz} we show the constraints on all of the 
parameters where $\Ptt$ helps significantly, not just dark 
energy.
This figure is for the specific case with Planck, Stage II, BAO from the same 
galaxy 
survey included,  $k_{\rm max}(z)=
0.1~\left[D\left(z\right)/D\left(0\right)\right]^{-1}\ihmpc$, 
and S/N=10
(note that, with one exception that we will indicate, $k_{\rm max}$ generally
does not apply to the BAO calculation, which essentially follows the          
procedure in \cite{2007ApJ...665...14S}). 
We see that dramatic improvements will be made in the constraints on 
$\Omega_k$, $\Omega_\Lambda$ (the dark energy density), and even $\omega_m$
and $A$.

\subsection{Specific example surveys}

We now explore a few planned or existing surveys.

\subsubsection{High redshift, high volume space mission}

\begin{figure}
\subfigure{\includegraphics[width=0.49\textwidth]{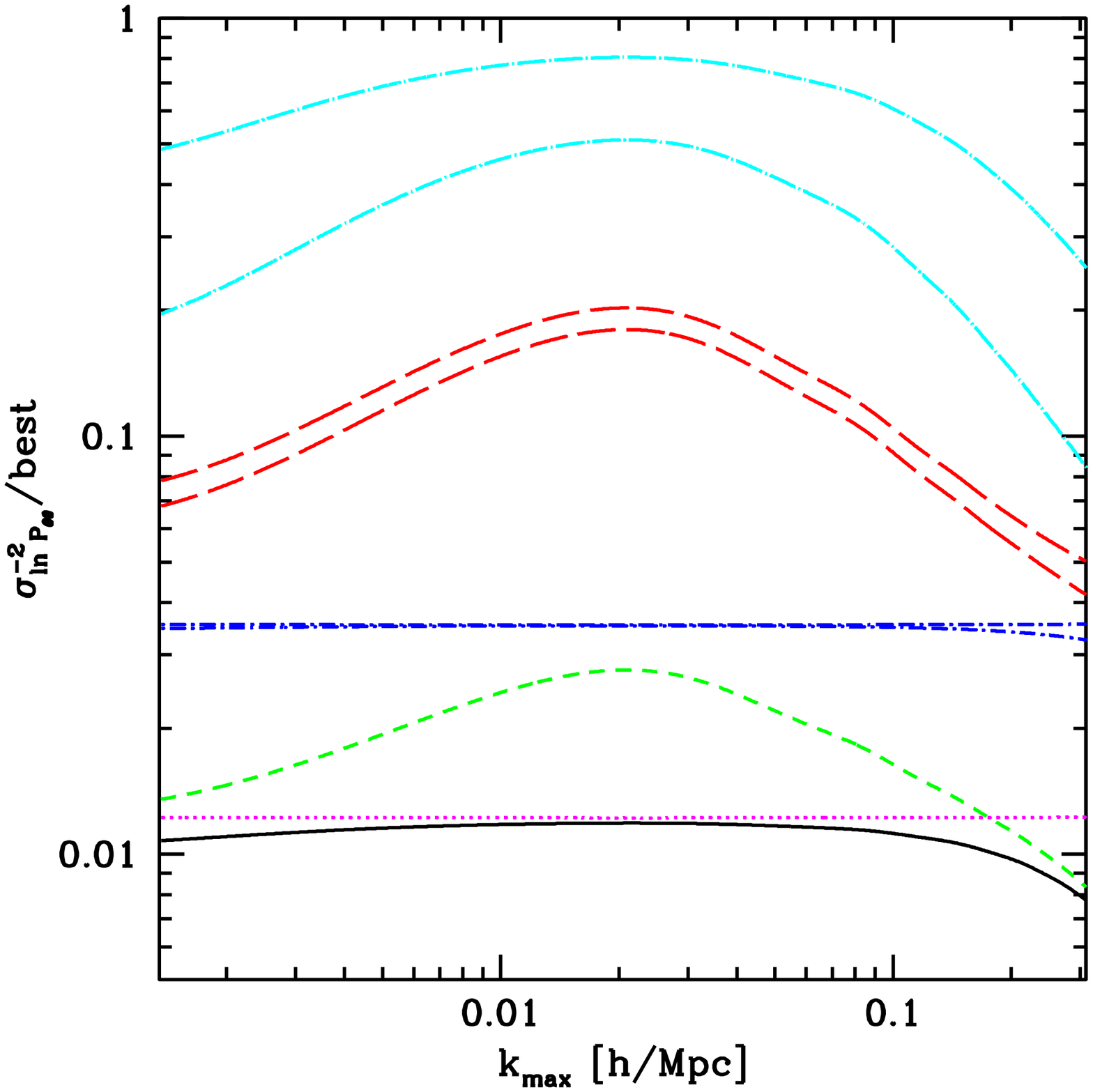}}
\subfigure{\includegraphics[width=0.49\textwidth]{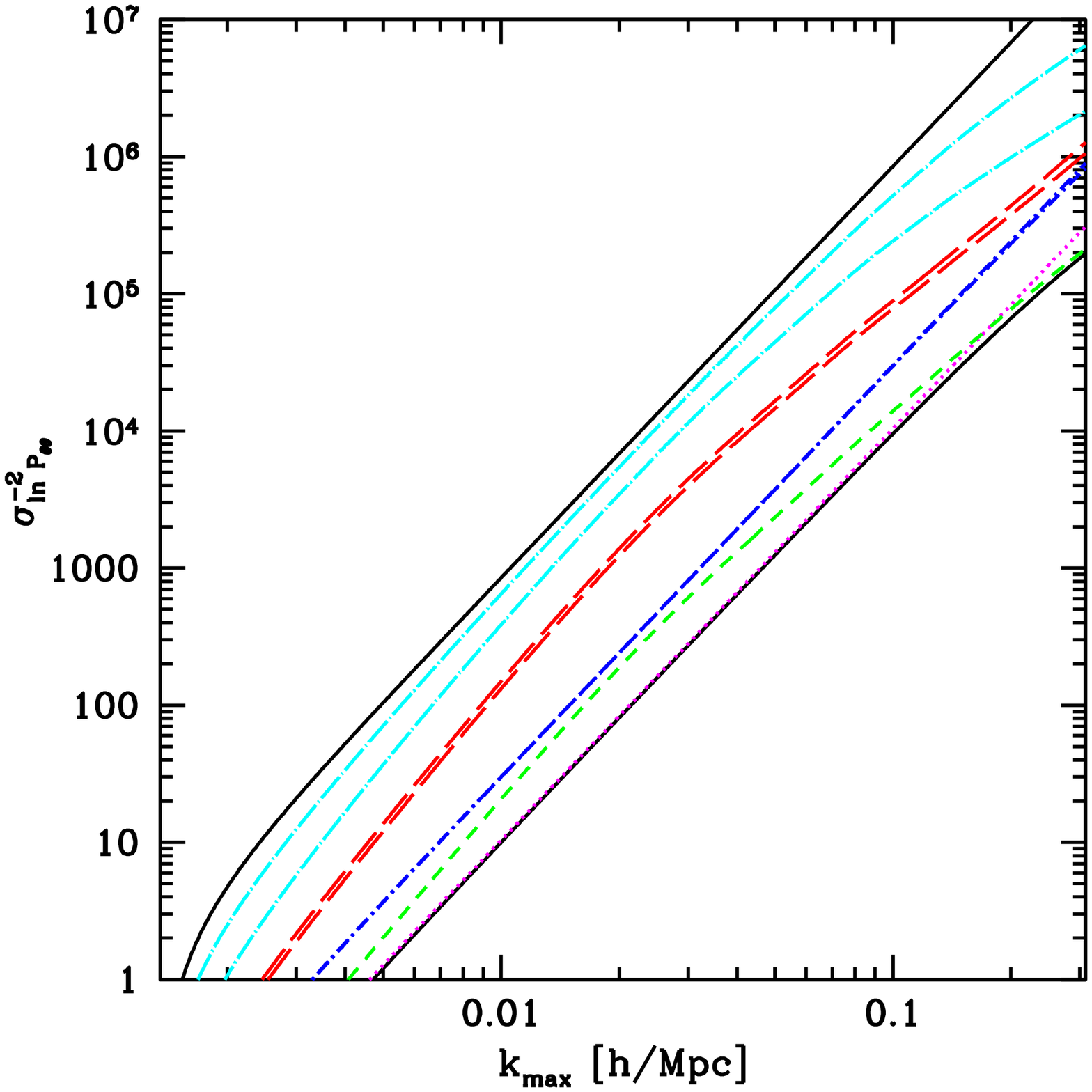}}
\subfigure{\includegraphics[width=0.49\textwidth]{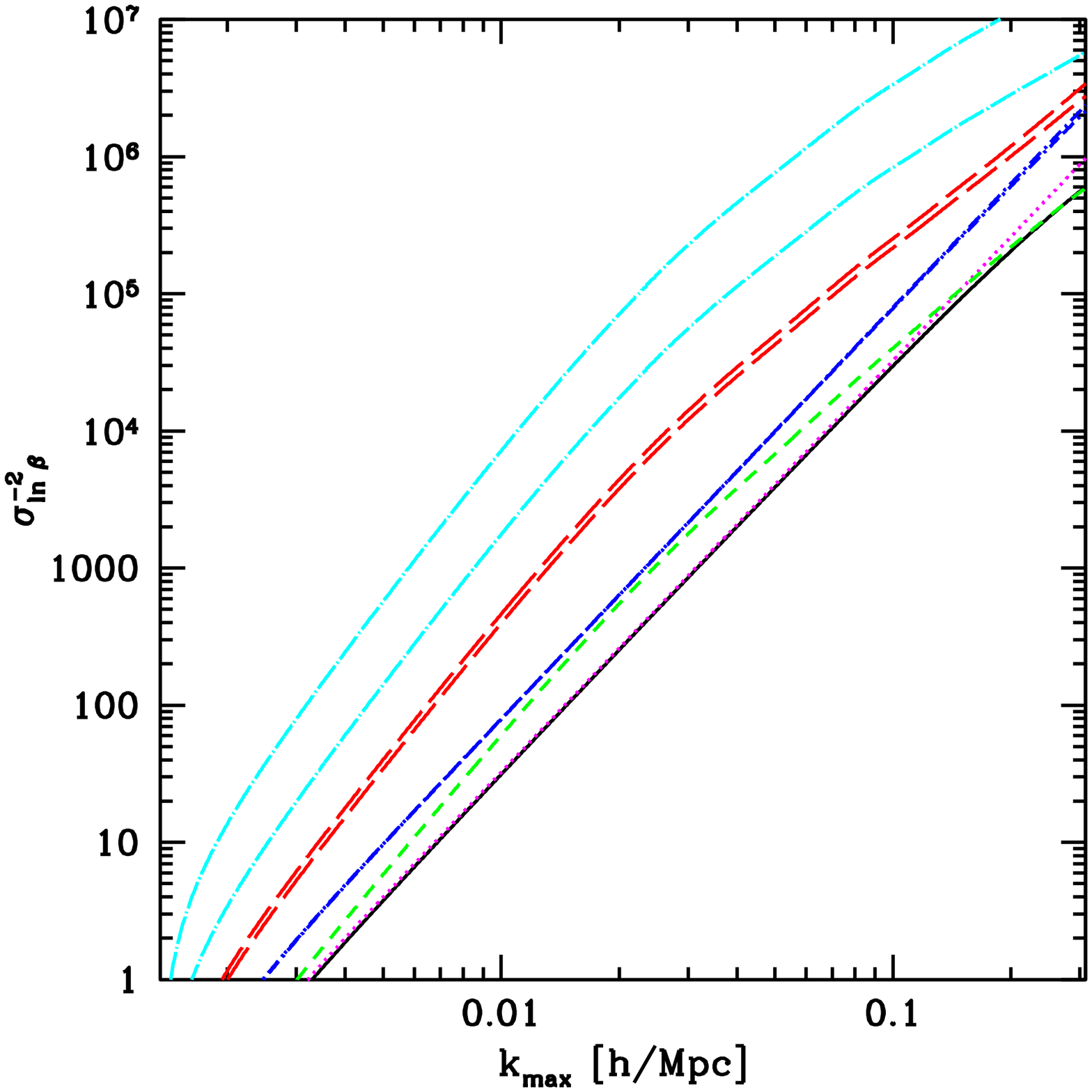}}
\subfigure{\includegraphics[width=0.49\textwidth]{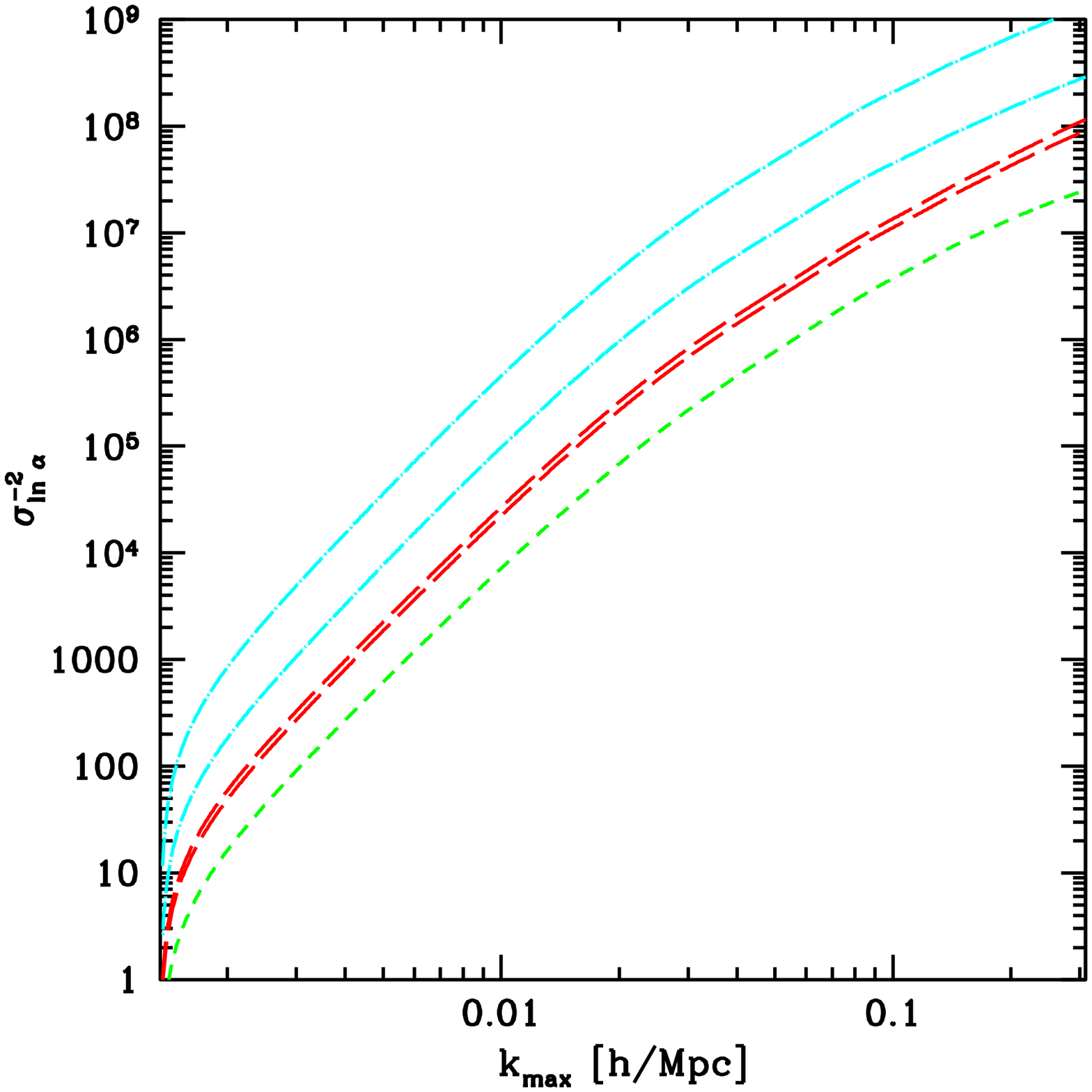}}
\caption{
Upper left panel:
Projected error (inverse variance) on the normalization of 
$\Ptt\equiv f^2 P_m$, relative to the error one could achieve by
observing velocity divergence 
directly (which is also obtainable by observing two noise-free
tracers with different bias), as a function of $k_{\rm max}$, for 
the EUCLID mission-motivated example described in the text.  
The black (solid) line shows EUCLID-like galaxies treated as a single tracer.
The magenta (dotted) and blue (dot-short-dashed) lines show single
perfectly sampled tracers with $b=2$ and $b=1$, respectively. 
Green (short-dashed) shows the EUCLID-like galaxies split into two groups with
different bias, as described in the text.  Red (long-dashed) shows the
EUCLID-like galaxies plus a perfect unbiased tracer, while cyan 
(dot-long-dashed)
shows a density $\bar{n}=0.03 \left(\hmpc\right)^{-3}$ plus an unbiased tracer. 
Where there are two lines of the same type, the upper one uses a genuinely
perfectly sampled tracer, while the lower one shows density 
$\bar{n}=0.03 \left(\hmpc\right)^{-3}$. 
Because the error is always dominated by $k$'s near $k_{\rm max}$, this 
plot (and subsequent plots) also gives a good estimate of how well one 
can do near $k_{\rm max}$ in a scale dependent measurement.
Upper right panel: Projected absolute error on $\Ptt$ for the same cases, 
except the upper black (solid) 
curve is for perfectly observed velocity divergence (or two perfect tracers
with different bias). 
Lower left (right) panel: Projected error on $\beta$ ($\alpha$), which has no
cosmic variance limit. 
}
\label{figeuclid}
\end{figure}

As a first example, we consider scenarios motivated by the EUCLID mission 
\cite{2008ExA...tmp...12C} (EUCLID is the combination of the missions formerly
known as SPACE and DUNE -- there is a proposal similar to SPACE in the USA 
named ADEPT). 
The proposal is to cover $\sim 1/2$ of the sky over a 
wide range of redshifts.  Our calculation considers  
$100 \left(h^{-1}{\rm Gpc}\right)^3$ centered at $z=1.4$.  We use the planned
galaxy density $0.0016 \left(\hmpc\right)^{-3}$
(L. Guzzo, private communication).
We assume a
mean bias of $b=2$ (somewhat arbitrarily -- the bias could be significantly 
higher if one is able to select galaxies in the most massive halos at that 
time).  Fig. \ref{figeuclid} shows the 
$f^2 P_m$ measurement for various modifications or additions to the EUCLID
mission.
We see (black, solid line, in the upper left) that this survey is a long
way from achieving the best possible measurement of $\Ptt$, by
almost a factor of 10 in the rms error.
Viewed as a single tracer, the problem is
not noise (compare to the magenta, dotted line), 
it is simply that, for $b=2$, we lose this much constraining power
to degeneracy with the bias.
A single perfectly sampled
tracer with $b=1$ would do better, but still be far from as good as 
possible (blue, dot-dashed). 
If we hypothetically split the
sample into $b=1.5$, $\bar{n}=0.0008 \left(\hmpc\right)^{-3}$ and
$b=2.5$, $\bar{n}=0.0008 \left(\hmpc\right)^{-3}$ subsamples (chosen for 
illustration, not because we know this is possible), in order to exploit
the method of \cite{2008arXiv0807.1770S}, 
the gains are non-negligible, especially
on large scales where they can more than double the effective volume of the 
survey, however, the noise is too large to achieve the much larger
gains that are in principle possible.
If we could leave the EUCLID-like sample intact and add a 
perfect unbiased tracer as the second field, we would get a substantial gain
from the combination, equivalent to multiplying the volume of the survey by a
factor of $\sim 10-17$
(the exact improvement factor depends on $k_{\rm max}$), although a factor 
$\sim 3$ of this would be obtained from the perfect unbiased tracer alone. 
As a practical matter,
a factor of ten larger number density for the unbiased tracer over the biased
tracer is nearly 
equivalent to perfect, i.e., the rms error is within $\sim 10$\% of the  
infinite density error.

The ratios shown in the upper left of 
Fig. \ref{figeuclid} are largely independent of the volume of the
survey.  The upper right of Fig. \ref{figeuclid} shows how well one can do in 
an absolute sense with a 100 cubic Gpc/h survey. 
One critical point to take away from this figure is how vital it is to push
$k_{\rm max}$ to be as large as possible. The overall error on a power spectrum
scales like $k_{\rm max}^3$ as long as one has good S/N, so a mere 30\%
slippage in $k_{\rm max}$ (e.g., $0.1 \ihmpc$ instead of $0.13\ihmpc$)  is 
equivalent to throwing away more than half of the survey volume! Similarly,
when one is using the multi-tracer method in this paper, increasing  
$k_{\rm max}$ by a factor of two is roughly equivalent to an order of magnitude
increase in the number of galaxies. Future analyses will need to explore
$k_{\rm max}$ of the single versus double tracer method, but for this numerical 
simulations are needed, which are beyond the scope of this paper. 

Figure \ref{figeuclid} also shows the projected error on
$\beta$, which shows very similar dependences to $f^2 P_m$.  
In fact, the
$\beta$ and $f^2 P_m$ errors are almost perfectly correlated, to the point 
where
it is probably most accurate to think of these surveys as making a very 
precise
measurement of $b^2 P_m$ which is converted to $f^2 P_m$ by multiplying by 
$\beta^2$, with the resulting error dominated by the error on $\beta$. 
Note, however, that with very low noise, the
$\beta$ error will continue to decrease toward zero, while the $b^2 P_m$ error
will not and therefore will eventually dominate the $f^2 P_m$ error (of course,
this is just the cosmic variance limit for a direct measurement of velocity
divergence).
Fig. \ref{figeuclid} also shows the projected error on $\alpha$. 
$\alpha$ can be measured very precisely, e.g., even for the simple split of 
the EUCLID-like sample in half, $\alpha$ will be measured to $\sim 0.1$\%
for  $k_{\rm max}= 0.05\ihmpc$, 
and $\sim 1$\% for 
$k_{\rm max}\gtrsim 0.01 \ihmpc$.  
This will measure the scale dependence of $\alpha$
and thus allow a very precise test of
the onset of non-linear biasing.

\subsubsection{Higher density, smaller volume space mission}

\begin{figure}
\subfigure{\includegraphics[width=0.49\textwidth]{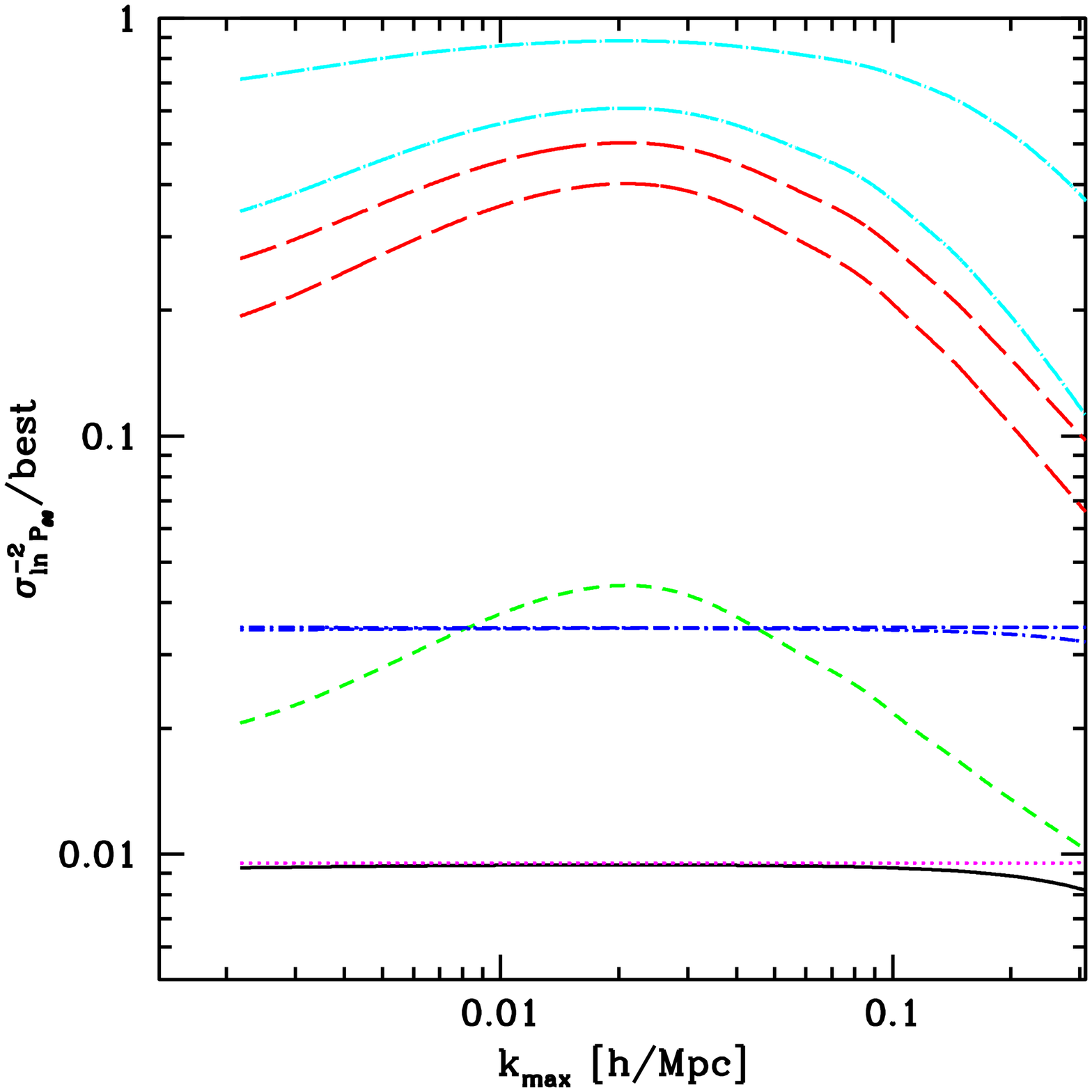}}
\subfigure{\includegraphics[width=0.49\textwidth]{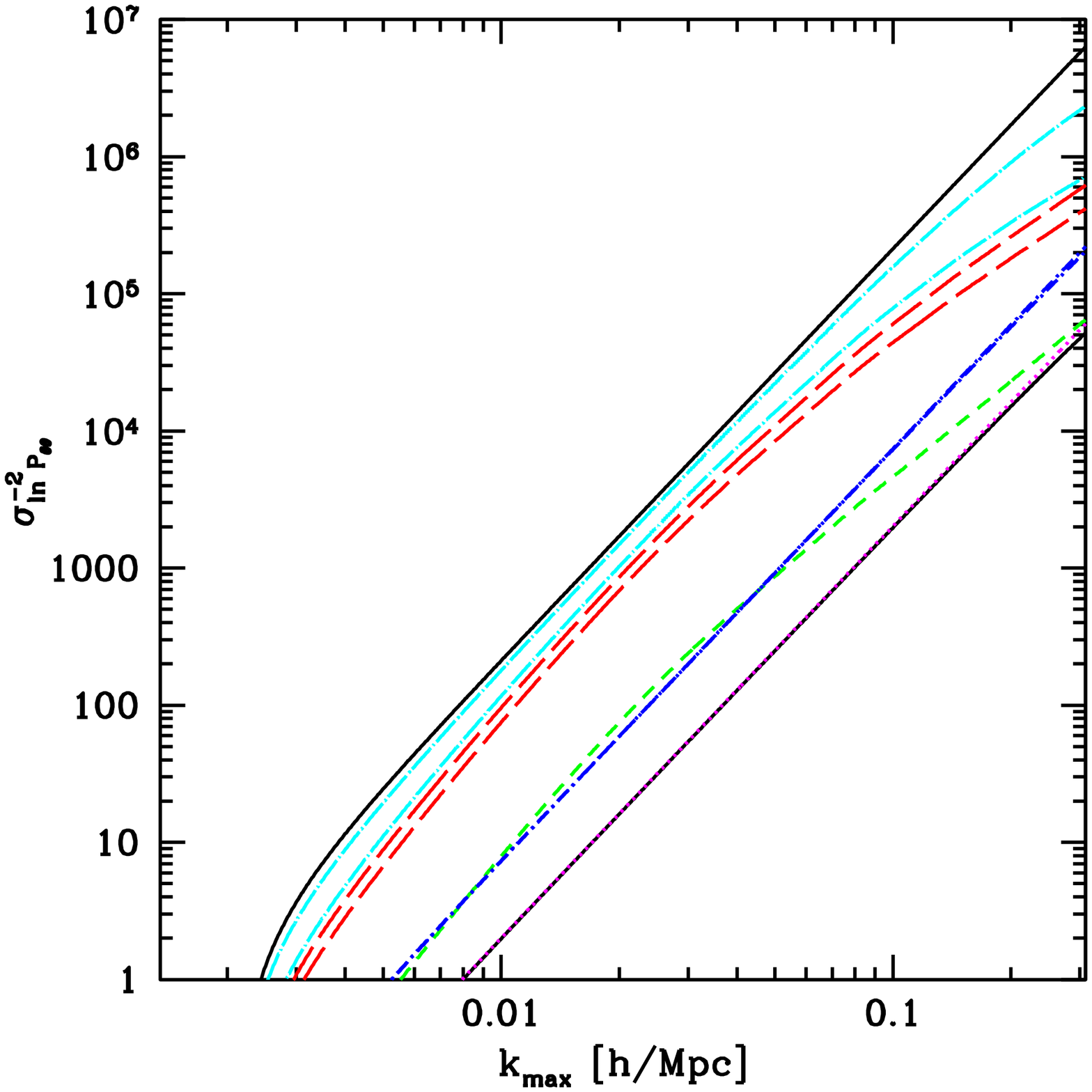}}
\subfigure{\includegraphics[width=0.49\textwidth]{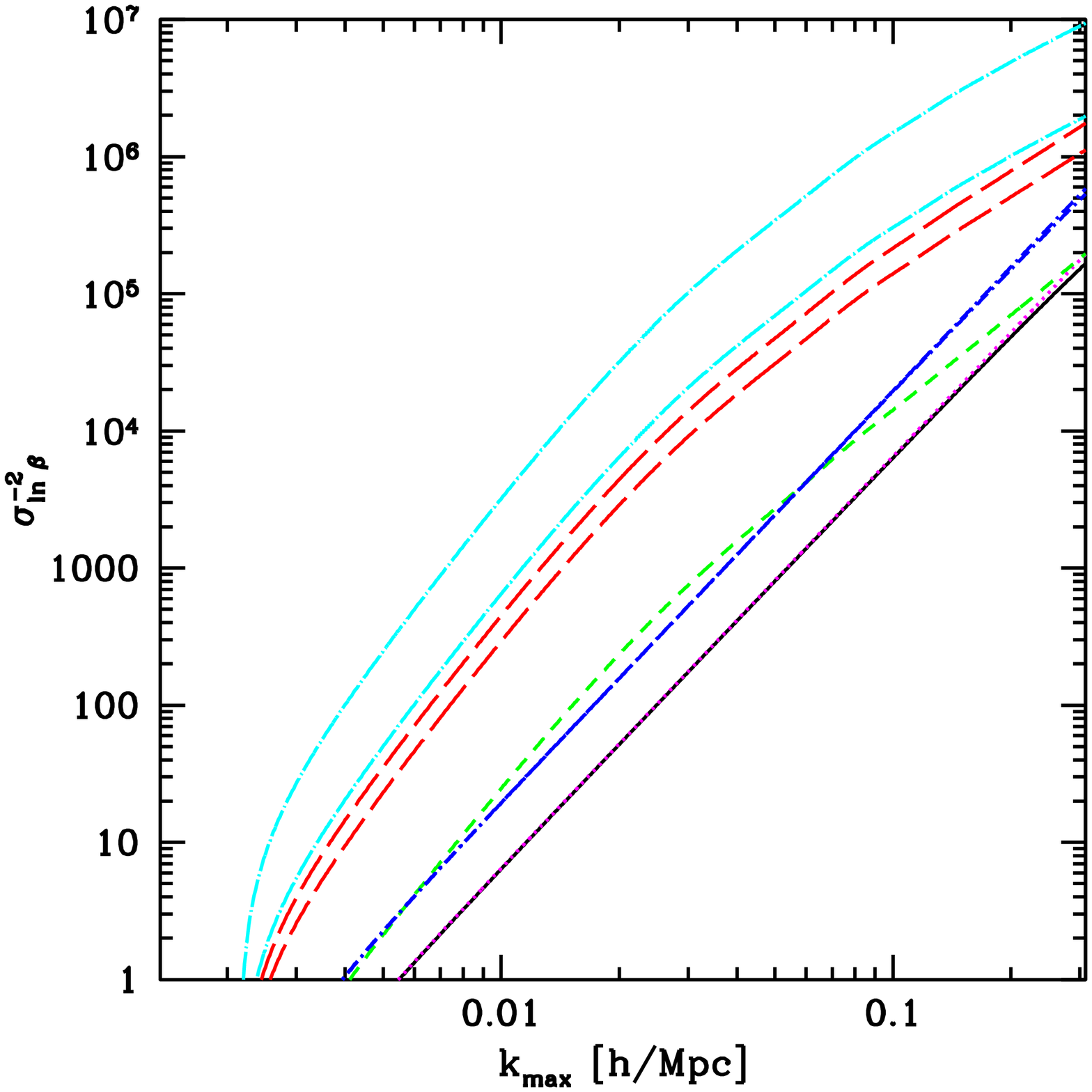}}
\subfigure{\includegraphics[width=0.49\textwidth]{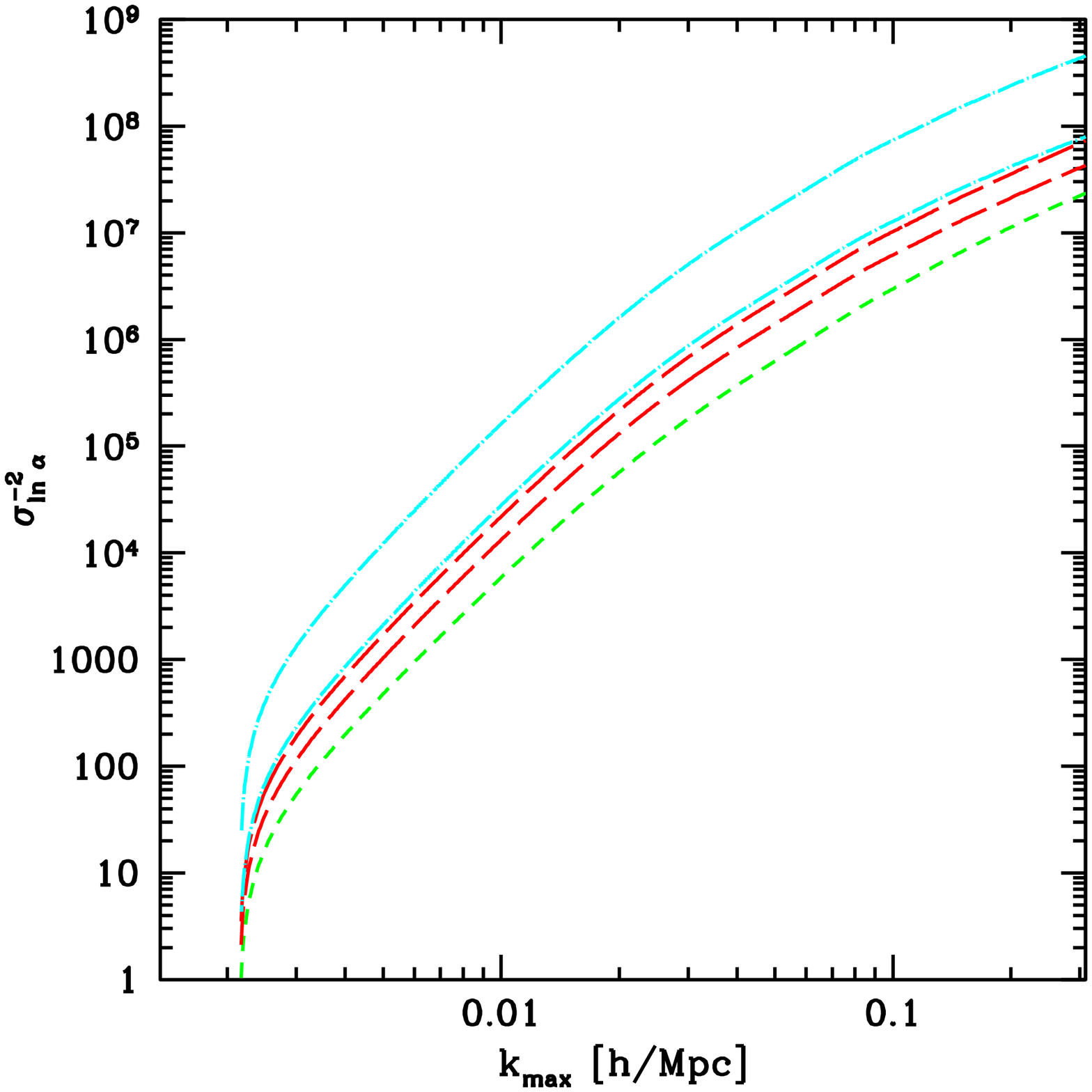}}
\caption{
Upper left panel:
Projected error (inverse variance) on the normalization of 
$\Ptt\equiv f^2 P_m$, relative to the error one could achieve by
observing velocity divergence 
directly (or using two perfect 
tracers with different bias), as a function of $k_{\rm max}$, for 
the SNAP mission-motivated example described in the text.  
The black (solid) line shows SNAP-like galaxies treated as a single tracer.
The magenta (dotted) and blue (dot-short-dashed) lines show single
perfectly sampled tracers with $b=2.3$ and $b=1$, respectively. 
Green (short-dashed) shows the SNAP-like galaxies split into two groups with
different bias, as described in the text.  Red (long-dashed) shows the
SNAP-like galaxies plus a perfect unbiased tracer, while cyan 
(dot-long-dashed)
shows a density $\bar{n}=0.03 \left(\hmpc\right)^{-3}$ plus an unbiased tracer. 
Where there are two lines of the same type, the upper one uses a genuinely
perfectly sampled tracer, while the lower one shows density 
$\bar{n}=0.03 \left(\hmpc\right)^{-3}$. 
The other panels are related to the upper left as in Fig. \ref{figeuclid}.
}
\label{figsnap}
\end{figure}

Another option to consider is a space mission with a
somewhat smaller volume, but higher density of galaxies.
For example, one of the SNAP satellite designs would allow for a 10000 sq. deg. 
survey at 
$0.9<z<1.6$, with density $\bar{n}=0.004 \left(\hmpc\right)^{-3}$ and mean
bias $b=2.3$ (D. Schlegel, private communication). The higher density than 
EUCLID is what we look for to make the
multiple-tracer method powerful. 
We show projections in Fig. \ref{figsnap} 
As an example, we split the SNAP sample into $b=2.8$ and $b=1.8$ subsamples.
The improvement in $\Ptt$ measurement is substantial, a factor of 
$\sim 2-5$ in effective volume, depending on $k_{\rm max}$. The enhanced
density of SNAP over EUCLID would roughly cancel the factor of 2 in greater
sky area in the EUCLID survey (note, however, that the EUCLID survey still 
comes out ahead overall because it covers a broader redshift extent).

\subsubsection{SDSS-III/BOSS}

\begin{figure}
\subfigure{\includegraphics[width=0.49\textwidth]{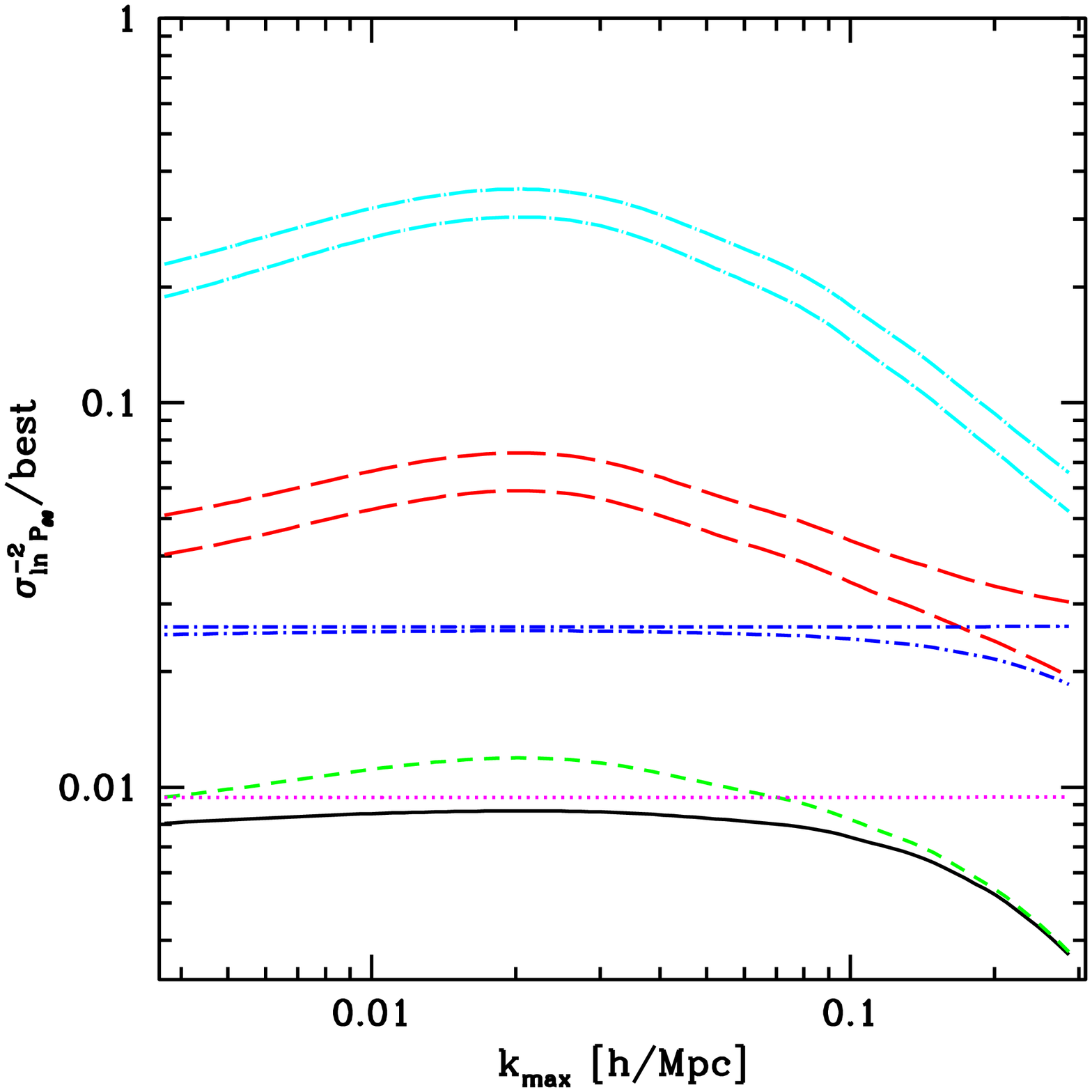}}
\subfigure{\includegraphics[width=0.49\textwidth]{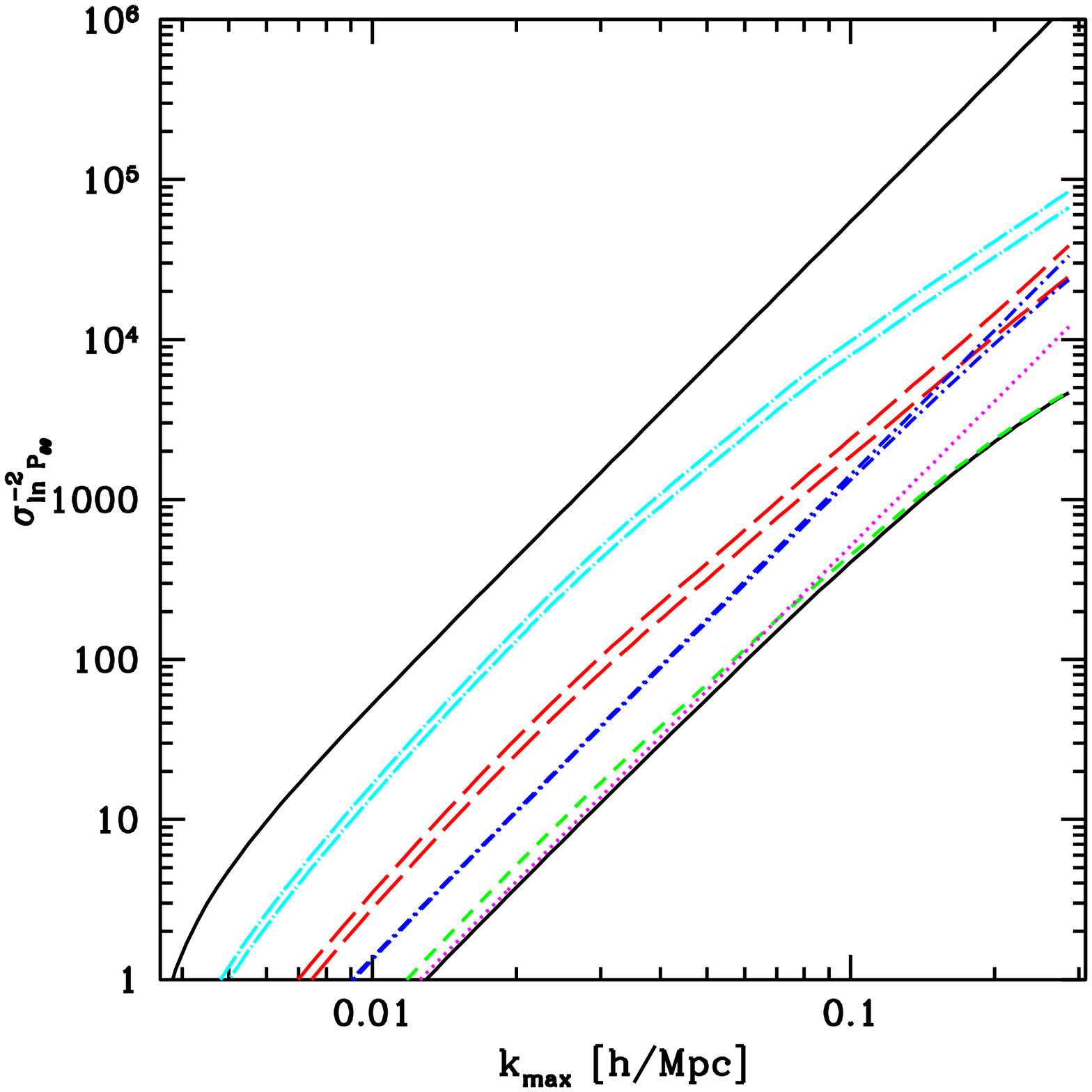}}
\subfigure{\includegraphics[width=0.49\textwidth]{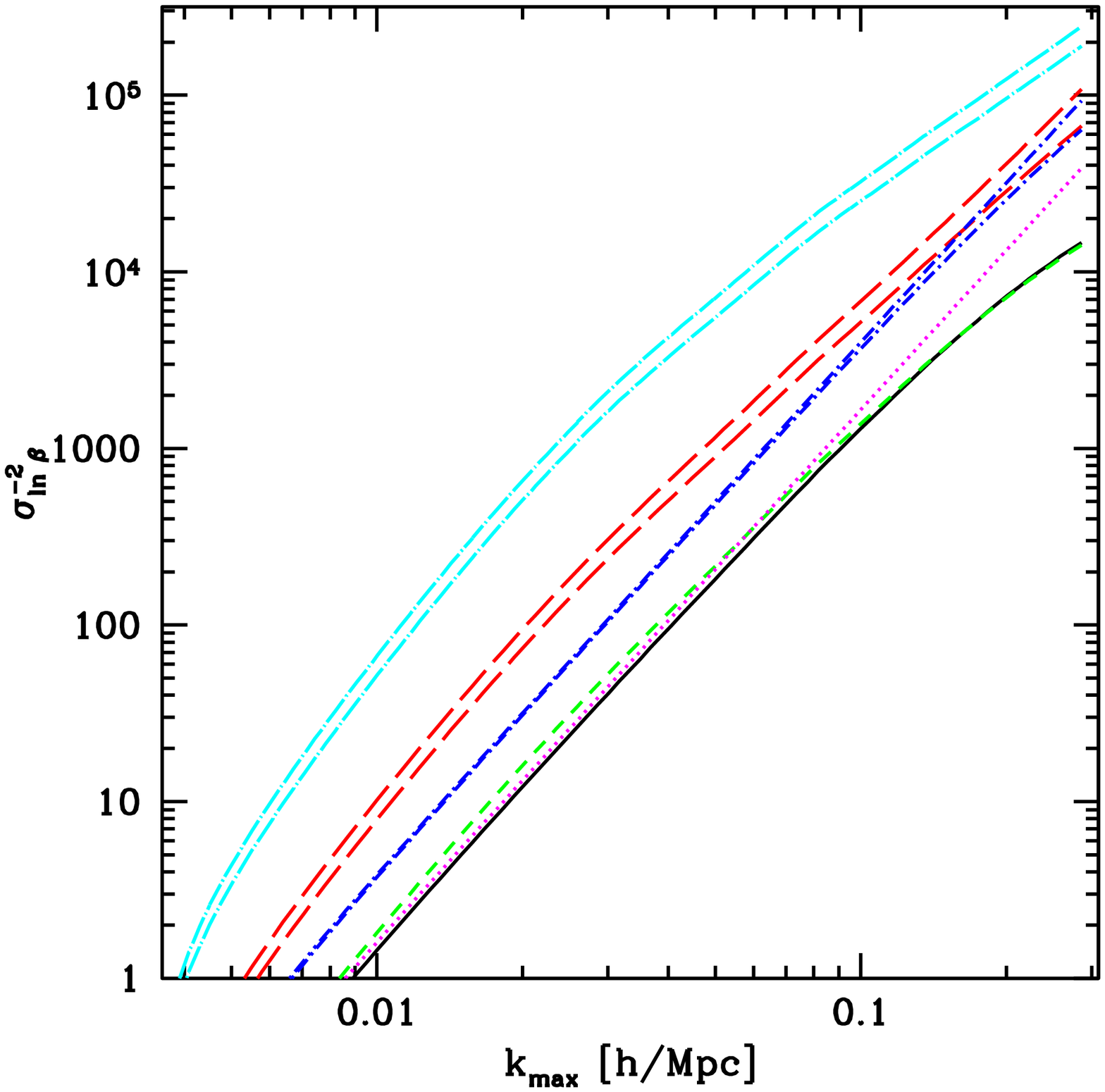}}
\subfigure{\includegraphics[width=0.49\textwidth]{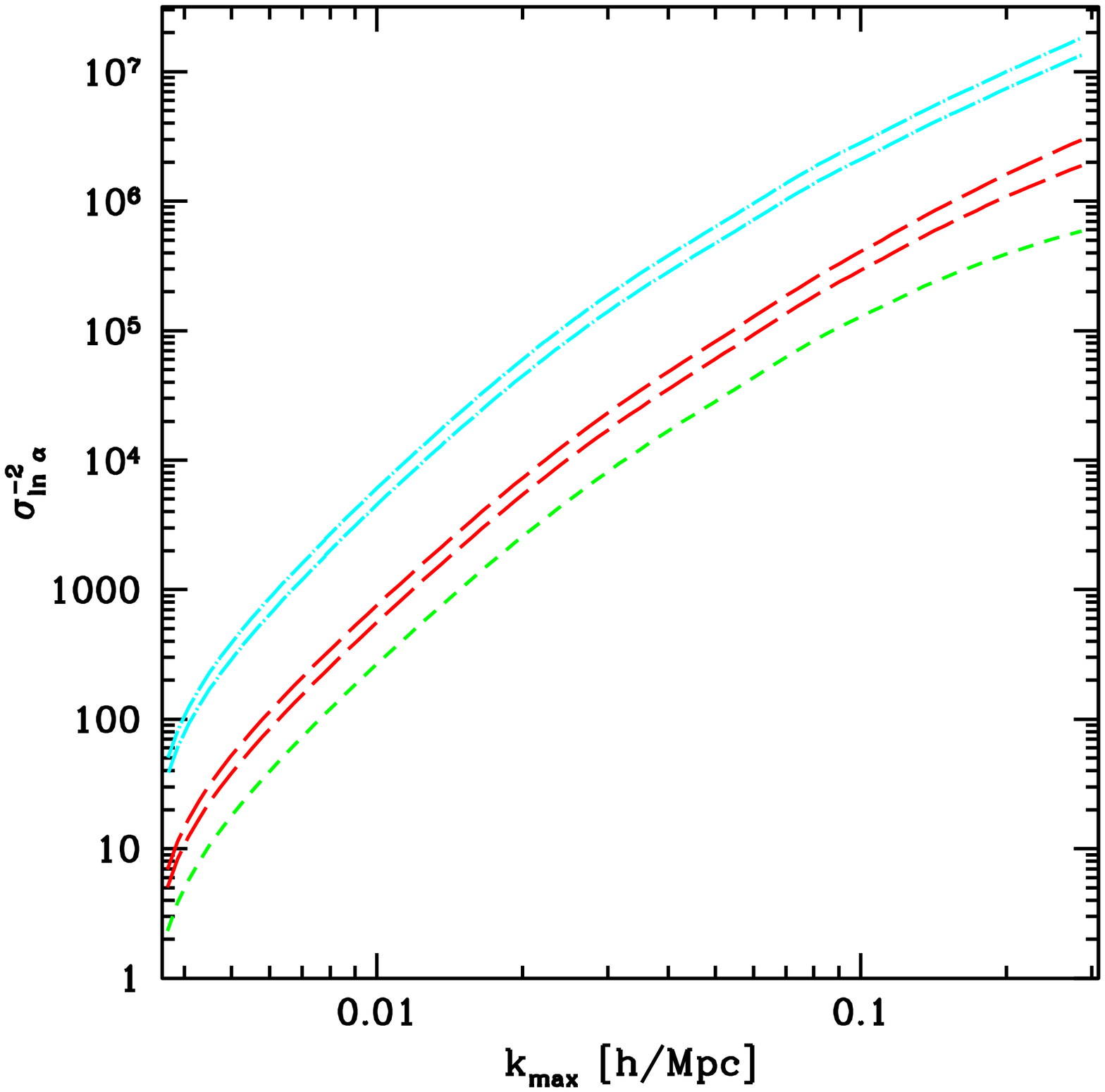}}
\caption{
Upper left panel:  Projected error (inverse variance) on the normalization of 
$\Ptt\equiv f^2 P_m$, relative to the error one could achieve by
observing velocity divergence
directly, or with two perfectly sampled tracers, 
similar to Fig. \ref{figeuclid},  
except this is for the SDSS-III/BOSS-galaxy-related example described in the 
text.  
The black (solid) line shows BOSS galaxies treated as a single tracer.
The magenta (dotted) and blue (dot-short-dashed) lines show single
perfect tracers with $b=1.9$ and $b=1$, respectively. 
Green (short-dashed) shows the BOSS galaxies split into two groups with
different bias, as described in the text.  Red (long-dashed) shows the
SDSS-III galaxies plus a perfect unbiased tracer, while cyan (dot-long-dashed)
shows ten times the BOSS galaxy density plus a perfect unbiased tracer.
Where there are two lines of the same type, the upper one uses a genuinely
perfectly sampled tracer, while the lower one shows density 
$\bar{n}=0.003 \left(\hmpc\right)^{-3}$ (long-dashed and dot-short-dashed) or
$\bar{n}=0.03 \left(\hmpc\right)^{-3}$ (dot-long-dashed). 
The other panels are related to the upper left as in Fig. \ref{figeuclid}.
}
\label{figlowz}
\end{figure}

Fig. \ref{figlowz} shows the achievable gains for survey configurations
motivated by the SDSS-III/BOSS survey \footnote{www.sdss3.org}.  
BOSS will cover 10000 sq. deg. with
central redshift $z=0.5$ and maximum $z\sim 0.7$, giving comoving volume  
6 cubic Gpc/h.  The planned galaxy sample has $b\sim 1.9$ and number density 
$\bar{n}=0.0003 \left(\hmpc\right)^{-3}$. 
Not coincidentally, the form of Fig. \ref{figlowz} is very similar to 
Fig. \ref{figeuclid}.  The only quantity that really matters for these 
calculations is the basic ratio of the noise level to power spectrum level, and
the optimization of these surveys for BAO means that this ratio will be very
similar. The BOSS S/N is somewhat lower than EUCLID, 
so BOSS is even 
farther from achieving the best possible measurement of $\Ptt$.
If we hypothetically split the
sample into a $b=1.6$, $\bar{n}=0.0002 \left(\hmpc\right)^{-3}$ and
$b=2.5$, $\bar{n}=0.0001 \left(\hmpc\right)^{-3}$ subsamples (chosen for 
illustration, not because this is known to be possible), in order to exploit
the method of \cite{2008arXiv0807.1770S}, the gains are small -- the noise is
too large. 
The different panels of Fig. \ref{figlowz} show how well one can do in an 
absolute sense with a 6 cubic Gpc/h survey, on $\Ptt$, $\beta$, and 
$\alpha$. 

\subsubsection{Present SDSS}

\begin{figure}
\subfigure{\includegraphics[width=0.49\textwidth]{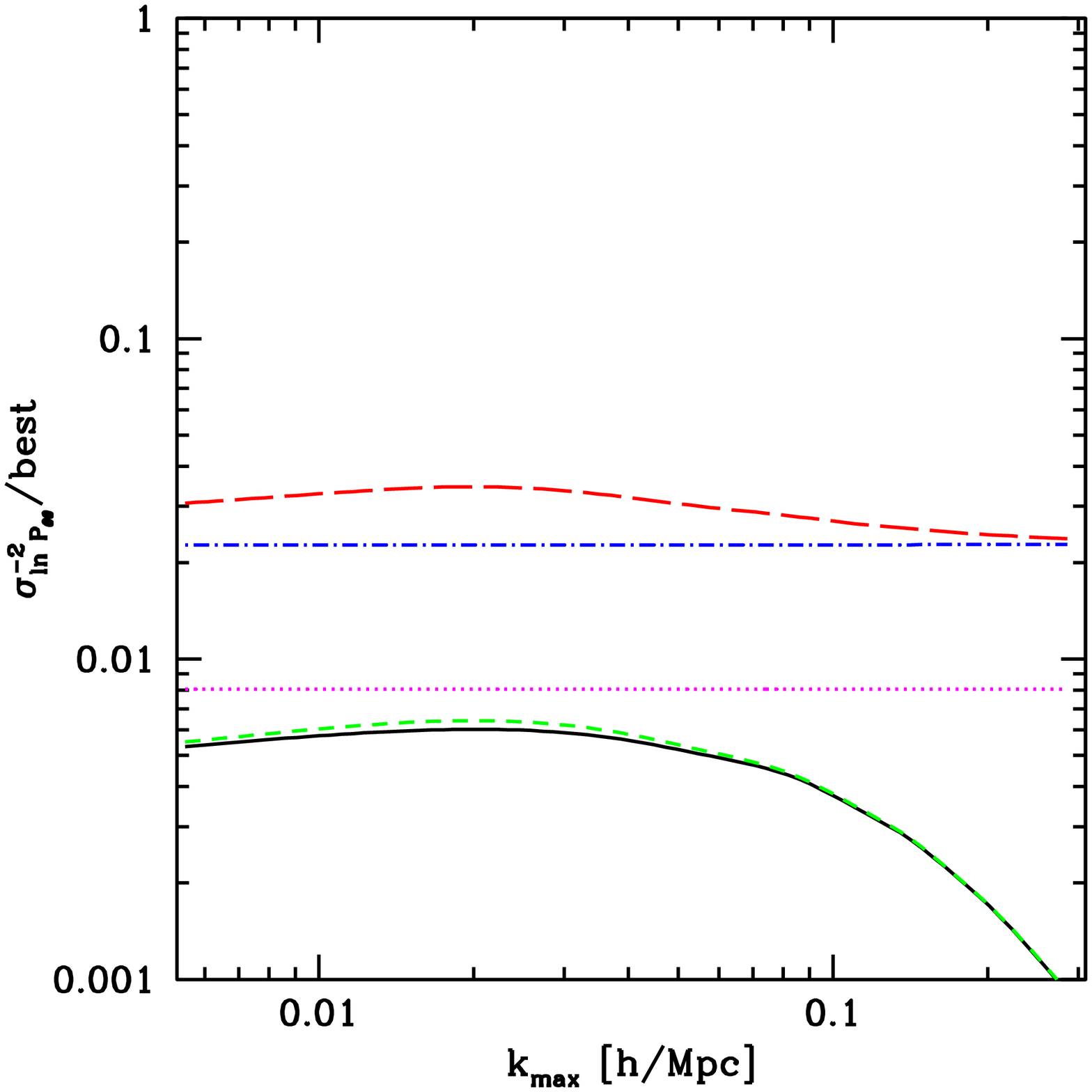}}
\subfigure{\includegraphics[width=0.49\textwidth]{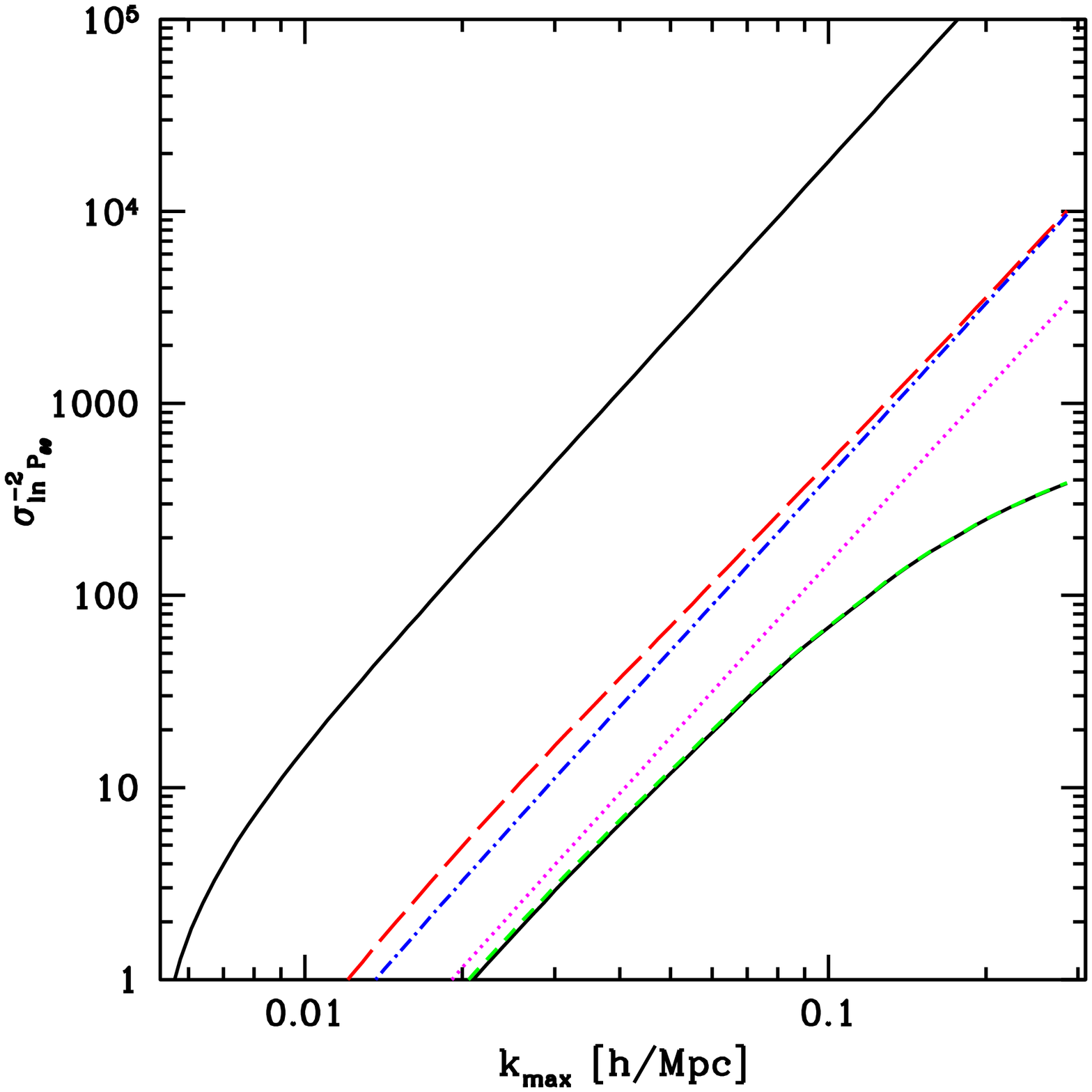}}
\subfigure{\includegraphics[width=0.49\textwidth]{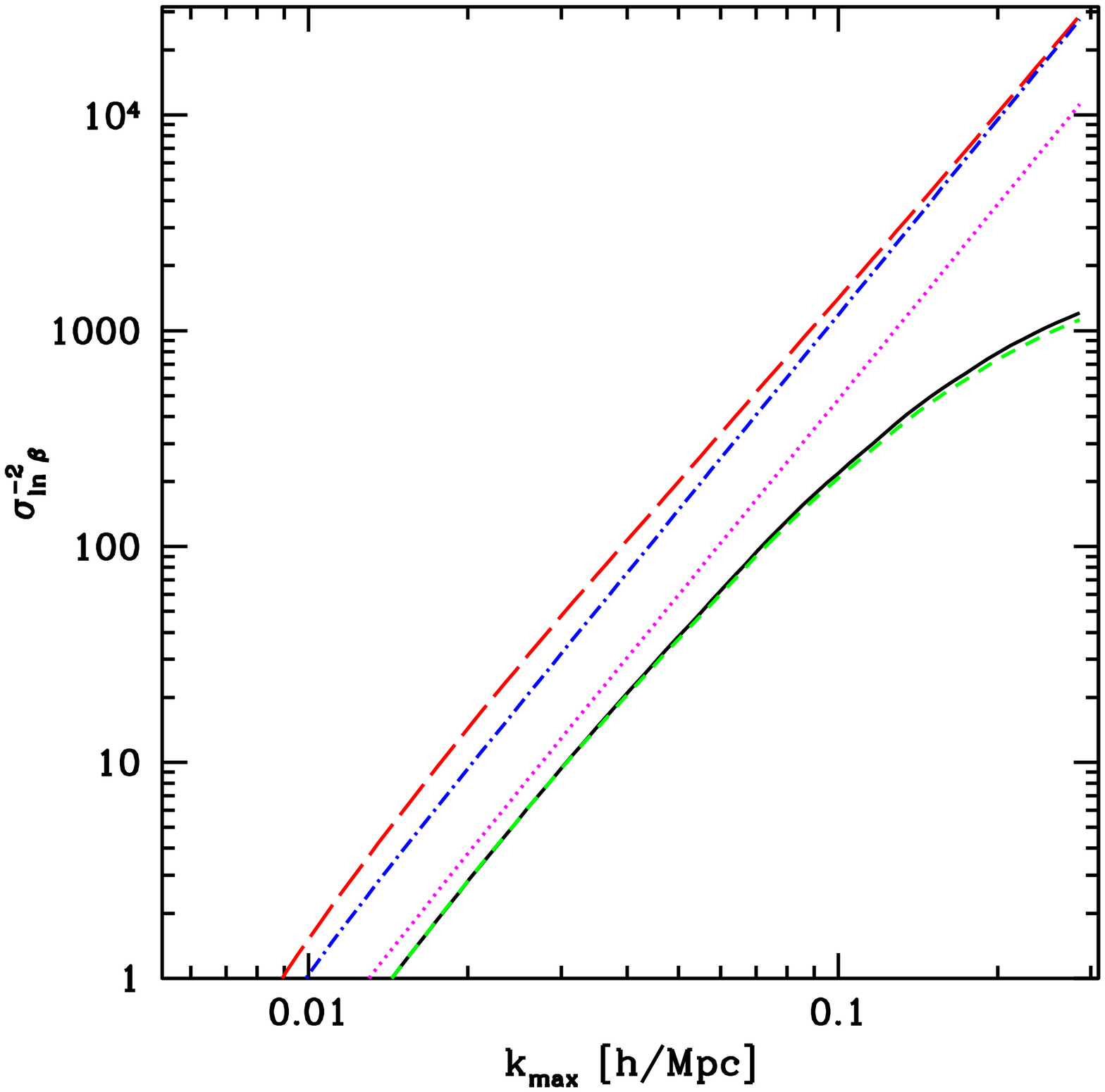}}
\subfigure{\includegraphics[width=0.49\textwidth]{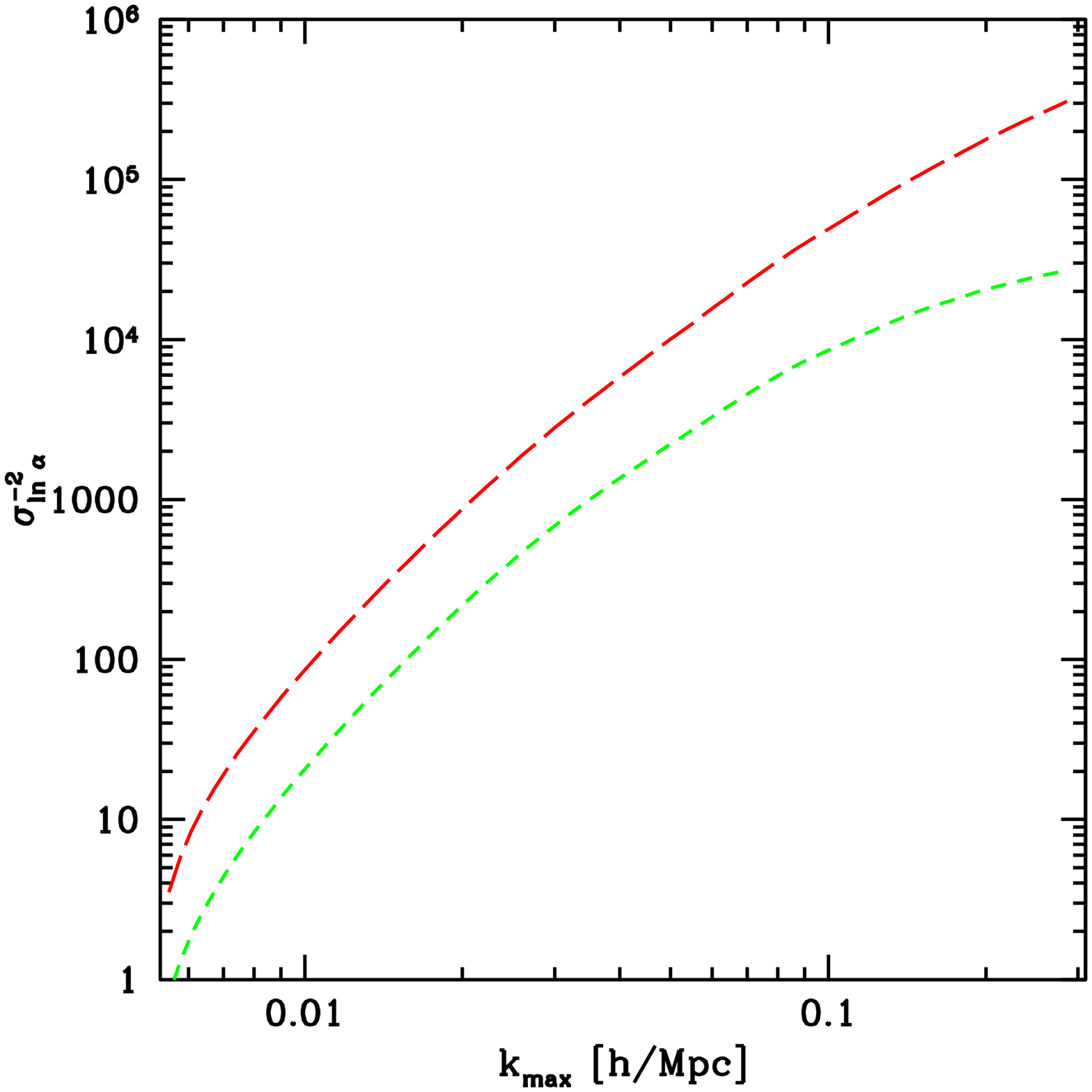}}
\caption{
Upper left panel:  Projected error (inverse variance) on the normalization of 
$\Ptt \equiv f^2 P_m$, relative to the error one could achieve by
observing velocity divergence
directly, or with two perfectly sampled tracers, 
similar to Fig. \ref{figeuclid},  
except this is for the current SDSS LRGs.
The black (solid) line shows LRGs as a single tracer.
The magenta (dotted) and blue (dot-short-dashed) lines show single
perfect tracers with $b=1.9$ and $b=1$, respectively. 
Green (short-dashed) shows the LRGs galaxies split into two groups with
different bias, as described in the text.  Red (long-dashed) shows the
LRGs plus a perfect unbiased tracer.
The other panels are related to the upper left as in Fig. \ref{figeuclid}.
}
\label{figlrg}
\end{figure}

To put the future in context, it is interesting to see what we could do with
current SDSS data. First, we take the LRGs, which we assume cover 
2 cubic Gpc/h centered at $z\sim 0.33$ (10000 sq. deg.) with a density 
$\bar{n}=0.000068\left(\hmpc\right)^{-3}$, and $b=1.9$.  
Fig. \ref{figlrg} shows our usual comparison between the $\Ptt$
measurement obtained from the LRGs and the ideal measurement.  
In this case, noise substantially degrades even the single-tracer version of
the measurement.  Splitting the sample into high and low bias parts 
(we assume $b=1.5$ and $b=2.3$) results in a completely negligible improvement,
and adding the LRGs on top of a perfect unbiased tracer does not improve the
results much (beyond the unbiased tracer results).
Figure \ref{figlrg} also shows the 
absolute results for $\Ptt$, $\beta$, and $\alpha$.  
While the
$\Ptt$ and $\beta$ figures are largely provided as a reference for
how well a single-tracer can do on these quantities, The lower right panel of
Fig. \ref{figlrg}
shows that the $\alpha$ measurement for the sample split in half could be
an interesting exercise. The bias ratio can be measured to about 2\% at 
$k\simeq 0.05\ihmpc$, or 1\% at $k\simeq 0.1\ihmpc$.  This should be precise
enough to place interesting constraints on the form of non-linear bias. 

\begin{figure}
\subfigure{\includegraphics[width=0.49\textwidth]{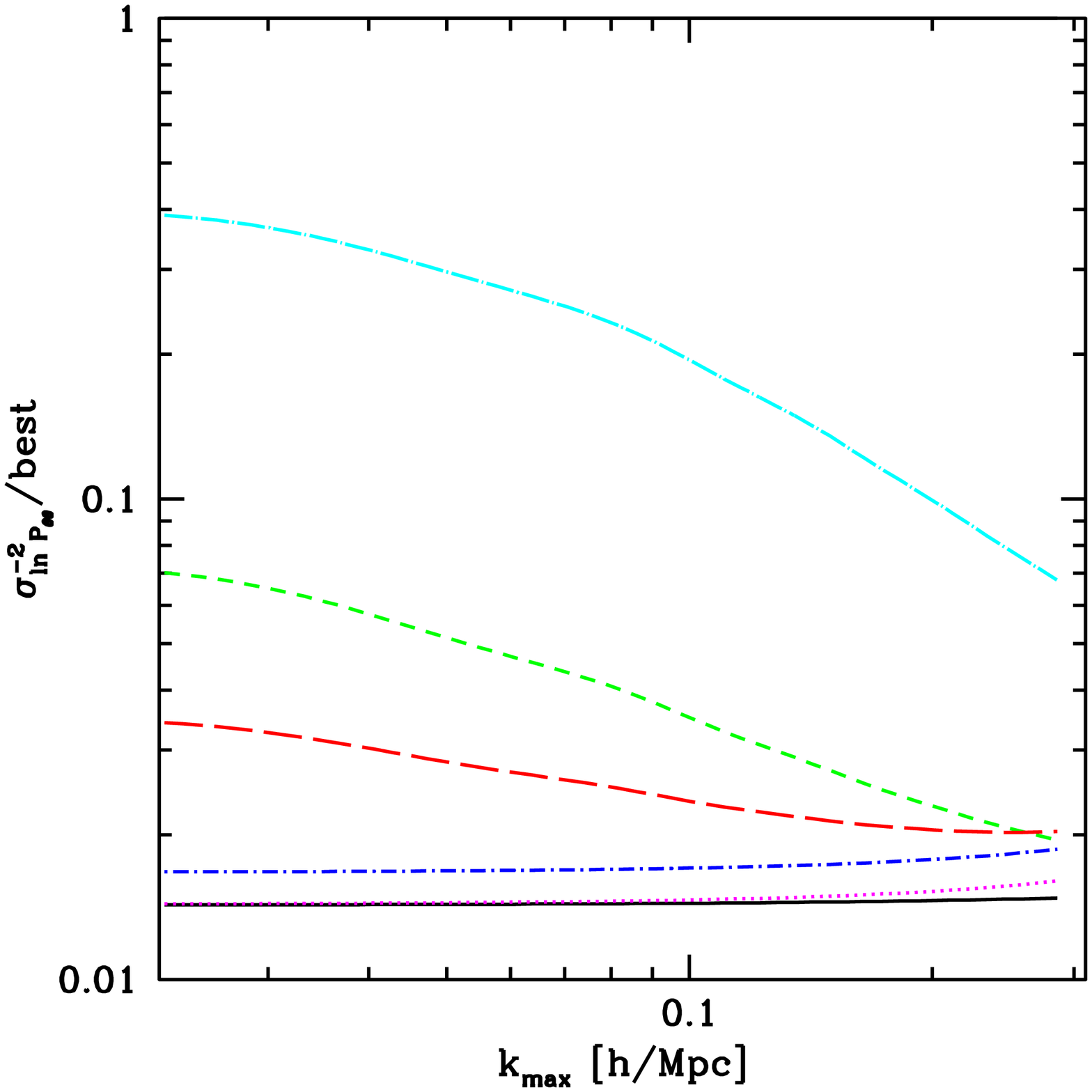}}
\subfigure{\includegraphics[width=0.49\textwidth]{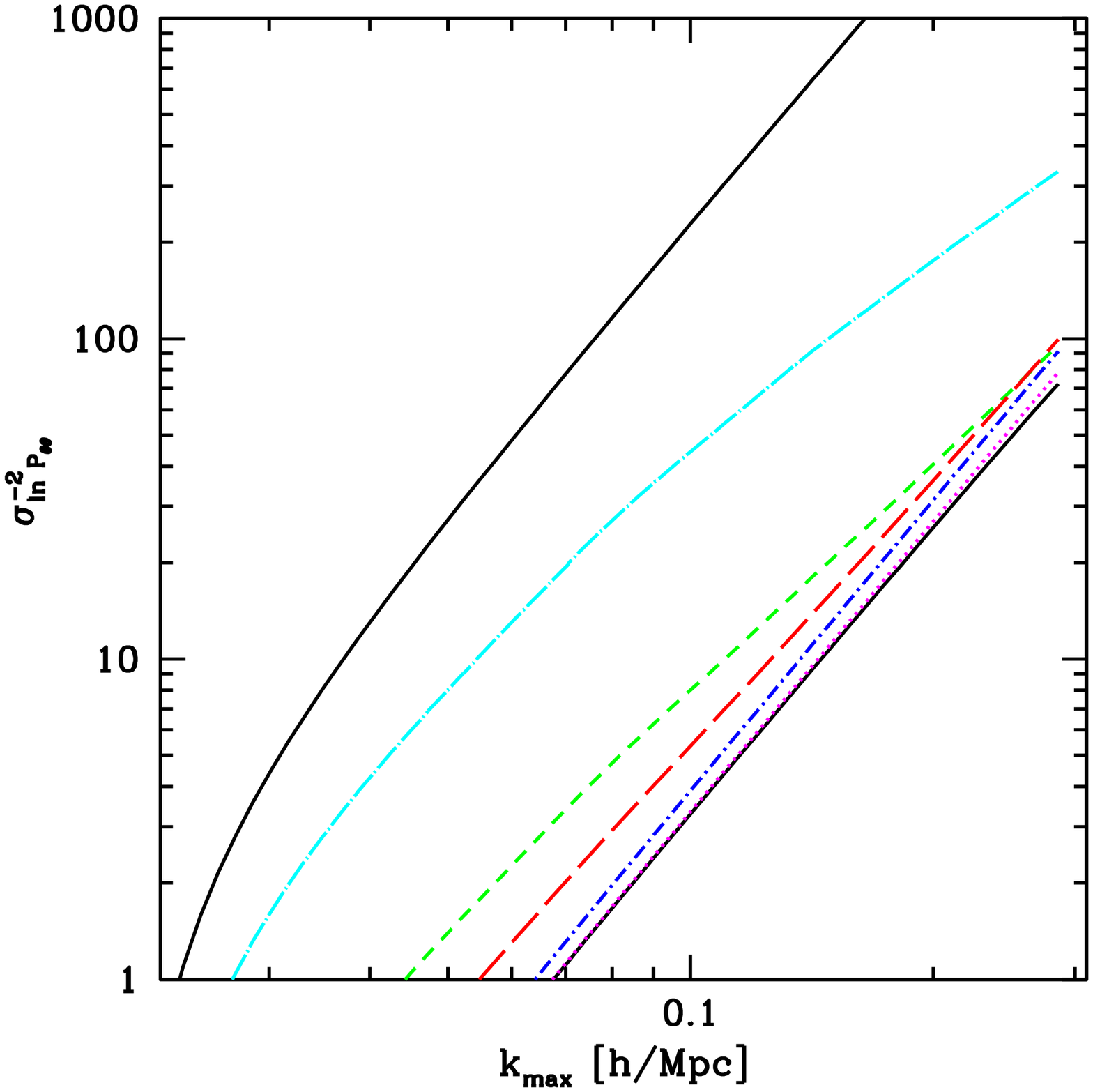}}
\subfigure{\includegraphics[width=0.49\textwidth]{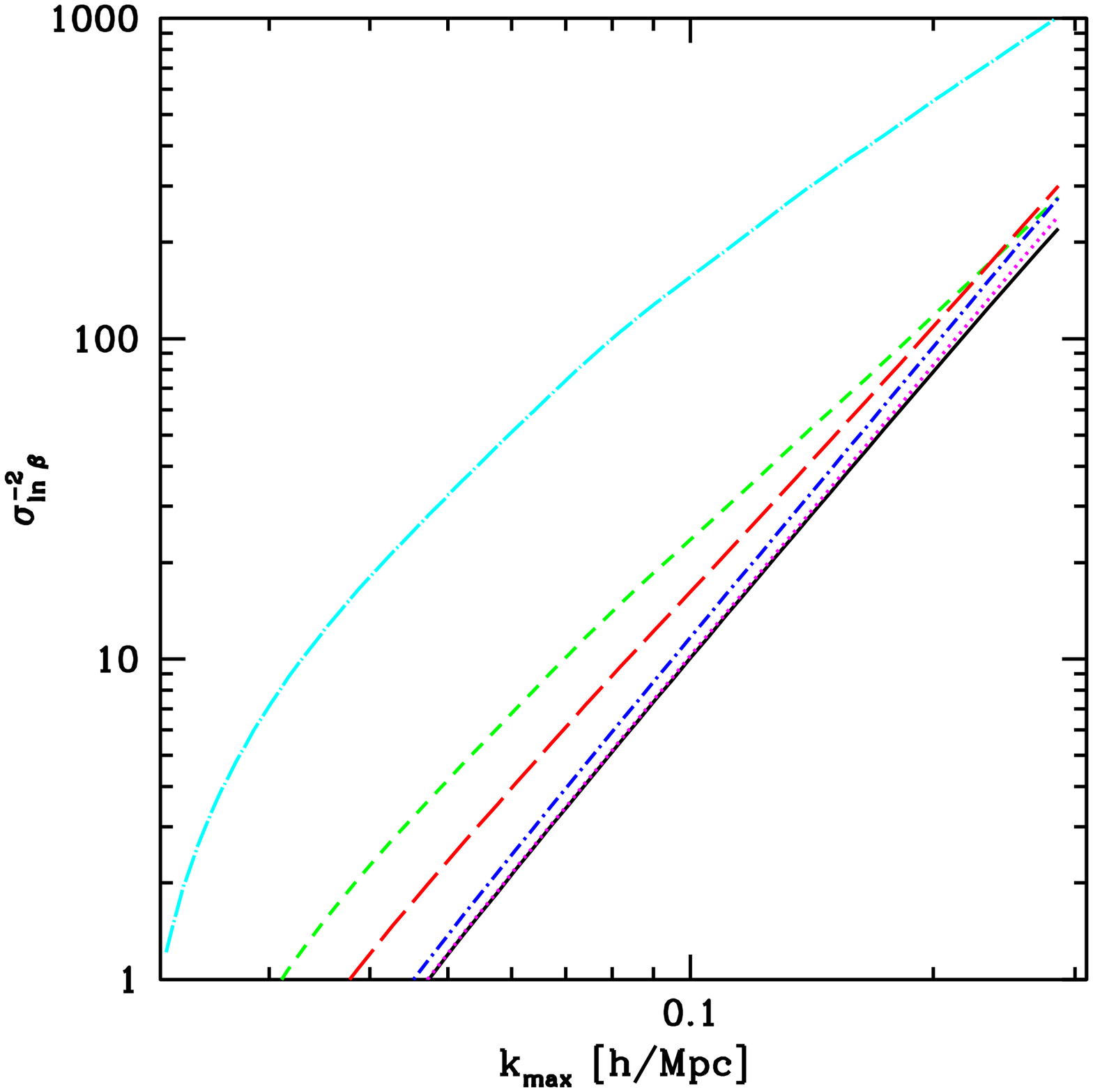}}
\subfigure{\includegraphics[width=0.49\textwidth]{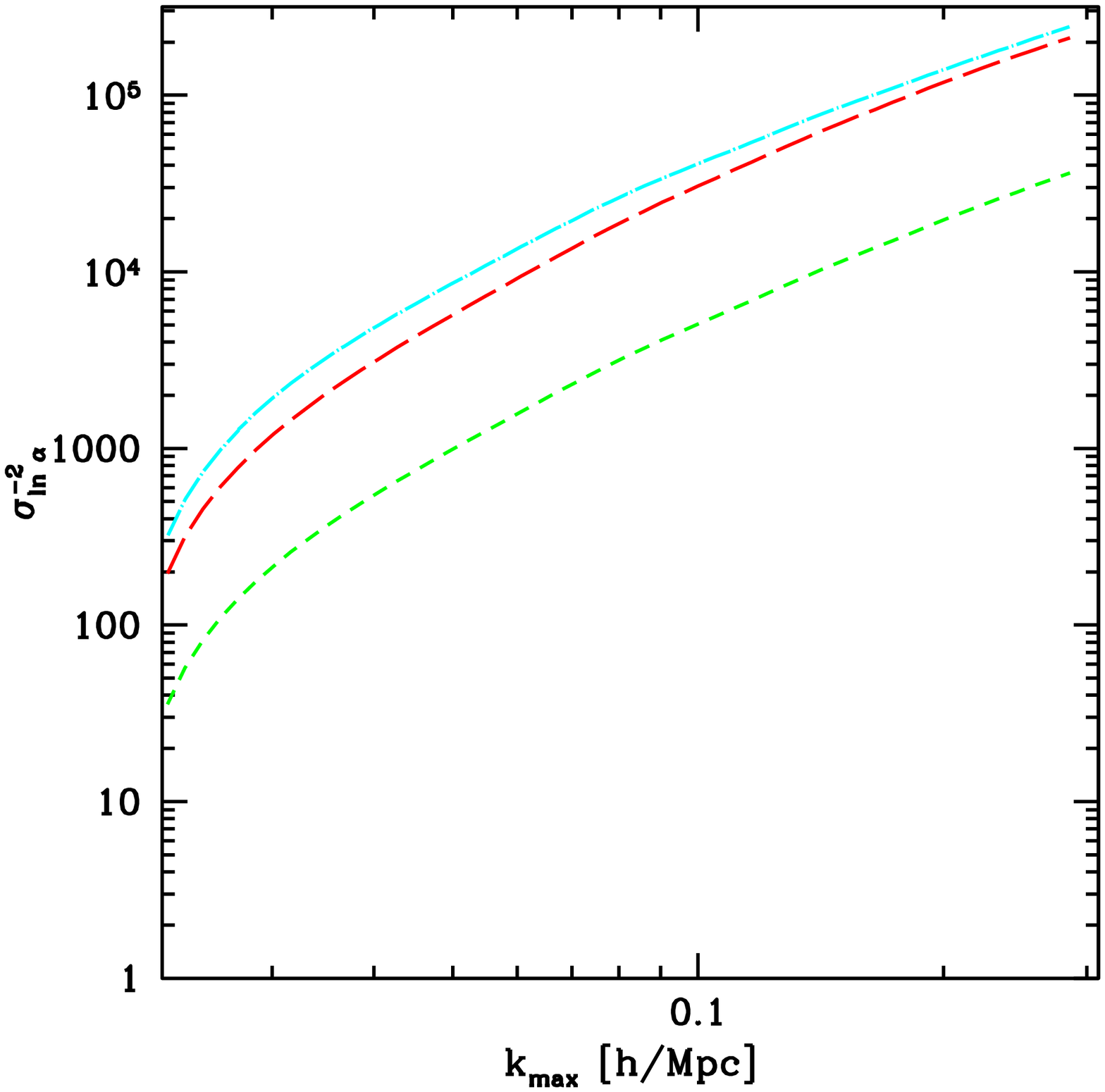}}
\caption{
Upper left panel:
Projected error (inverse variance) on the normalization of 
$\Ptt \equiv f^2 P_m$, relative to the error one could achieve by
observing velocity divergence
directly, or with two perfectly sampled tracers, 
similar to Fig. \ref{figeuclid},  
except this is for the current SDSS main galaxies.
The black (solid) line shows SDSS galaxies as a single tracer.
The magenta (dotted) and blue (dot-short-dashed) lines show single
perfect tracers with $b=1.1$ and $b=1$, respectively. 
Green (short-dashed) shows the main galaxies split into two groups with
bias $b=1.3$ and $b=0.9$, both with $\bar{n}=0.0044\left(\hmpc\right)^{-3}$, 
as described in the text.  Red (long-dashed) shows the
main galaxies plus a perfect unbiased tracer. 
Cyan (dot-long-dashed) shows two tracers
with $b=1.3$ and $b=0.9$, both with $\bar{n}=0.044\left(\hmpc\right)^{-3}$. 
The other panels are related to the upper left as in Fig. \ref{figeuclid}.
}
\label{figmain}
\end{figure}

The SDSS main galaxy sample is generally not as good for LSS as the LRGs, 
because it covers only low redshift and thus much less volume; however, it 
does have the higher number density that we look for to make the multi-tracer
method powerful. \cite{2008MNRAS.385.1635S} provide a realistic split into 
blue and red samples with bias ratio 0.7, each with 
$\bar{n}\simeq 0.0044 \left(\hmpc\right)^{-3}$ at $z<0.1$ (center of 
volume $z\simeq 0.08$).  We take $b=0.9$ and $b=1.3$. This sample covers only 
0.026 cubic Gpc/h. Fig. \ref{figmain} shows that the 
multi-tracer approach to measuring $\Ptt$ is quite effective at this
number density. 
Unfortunately, the upper right panel of Fig. \ref{figmain} shows that the 
absolute measurement still 
is not very good, requiring one to go to $k_{\rm max}>0.1$ just to get a 
$3-\sigma$ level of detection, and to $k_{\rm max}>0.3$ to measure 
$\Ptt$ to 10\% (which is surely not possible at low $z$ because of 
non-linearity).
The situation with $\beta$ is similar, but it is again interesting to note, 
in the lower right of Fig. \ref{figmain}, 
that the bias ratio for the realistic split sample
can be measured to a few percent. One should keep in mind however that 
at high $k$ the tracers are not maximally correlated and this weakens the 
constraints. Cosmological simulations will be needed to address this in more 
detail.

\subsection{Beyond redshift-space distortions}

\begin{figure}
\resizebox{\textwidth}{!}{\includegraphics{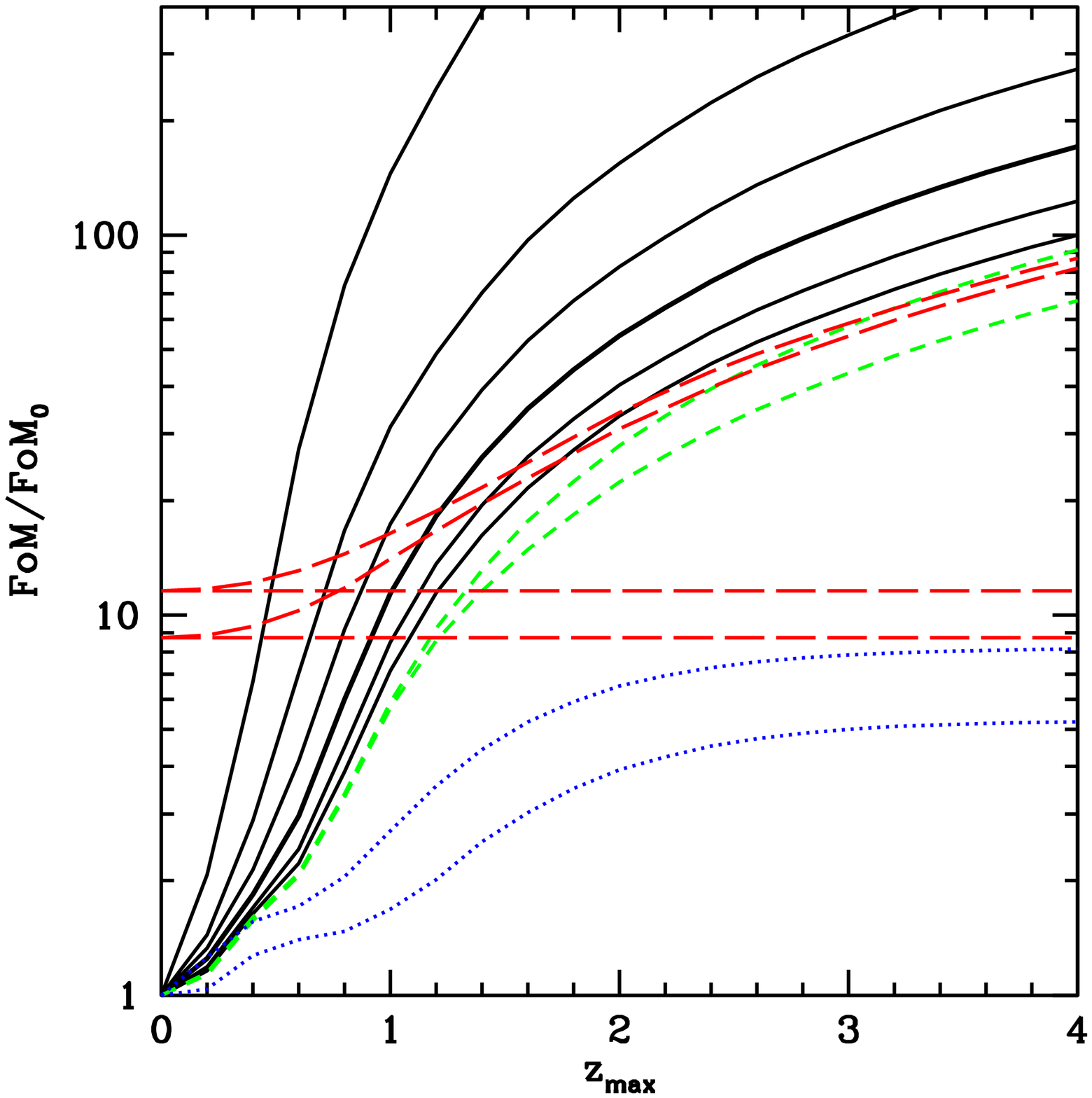}}
\caption{
Improvement in DETF FoM when all of the information in the redshift-space
power spectrum is used rather than just BAO and
$f^2 P_m$ constraints, for $k_{\rm max}(z)=
0.1~\left[D\left(z\right)/D\left(0\right)\right]^{-1}\ihmpc$ (relative to
Planck+Stage II, and for 30000 sq. deg., as usual).
Green (dashed) lines show single S/N=1
(at $k=0.4\ihmpc$) tracers with
$b=1$ (upper) and $b=2$ (lower). The black curves show both types of galaxy
together, each with (bottom to top) S/N=1, 3, 10, 30, 100, 1000 (the
constraining power increases without bound as the S/N improves).
Blue (dotted) shows BAO alone, with the lower curve using the above
$k_{\rm max}$ and
the upper curve using smaller scales as described in \cite{2007ApJ...665...14S}
(the rest of the curves effectively use the weaker version).
The red (long-dashed) curves include weak lensing constraints, with the lower
two (horizontal) lines including no redshift survey (lowest is approximately 
as SNAP weak lensing, 2nd lowest is LSST weak lensing), while the upper two include the $b=2$-only (weakest) redshift
survey.
}
\label{figfomvszAPwl}
\end{figure}

So far, primarily for the purpose of transparency of the presentation, the only
information we have used from galaxy surveys is the BAO scale and
$\Ptt(z,k=0.1\ihmpc)$, 
as measured from redshift-space distortions, ignoring
other effects on the galaxy power spectrum. Generally, however, several other
obvious effects should be included: 
The power spectrum cannot be observed
in $\hmpc$ units, so the conversion to velocity (redshift) coordinates,
$\Delta v = H(a)~a~\Delta x_\parallel$, and angular coordinates, 
$\Delta \theta = D_A(a)^{-1}~a~\Delta x_\perp$, must be accounted for. 
The uncertainty in the component of these conversions that takes one from 
angular coordinates to
velocity coordinates, $H(a) D_A(a)$, is often called the 
Alcock-Paczy{\'n}ski effect \cite{1979Natur.281..358A}.
In addition to the transformation to observable coordinates, cosmological 
parameters generally affect the scale dependence of the power spectrum through
the transfer function between early-time 
perturbations from inflation and the linear theory perturbations that we see,
e.g., the classic $\Gamma=\Omega_{m,0} h$ scale for the turnover of the 
power spectrum.
In Figure \ref{figfomvszAPwl} we show the constraining power for the
same general 3/4-sky survey up to $z_{\rm max}$ that we have discussed before, 
except
now computing the Fisher matrix using the full parameter dependence of the
power spectrum.
Note that the BAO measurement is automatically included within this framework,
although in a somewhat weaker than usual form because
of our relatively conservative $k_{\rm max}$ cut.
The extra parameter
dependence leads to typically a factor ~3-5 larger FoM than one would find
using $f^2 P_m$ and BAO constraints alone. There is a new feature
at high S/N, however,
in that we no longer find any cosmic variance limit on the constraining power!
The measurement can be improved arbitrarily much if the S/N can be improved, but 
to achieve these improvements the number density of tracers must greatly exceed current designs.

In Fig. \ref{figfomvszAPwl} we also explore the complementarity of weak
gravitational lensing measurements, as represented by the DETF
Stage IV space (SNAP) or ground (LSST) projections (we use their optimistic
versions). We see that these
experiments, in combination with Planck (and all Stage II experiments, although
these play a minor role), can achieve an FoM improvement of about a factor of
10 over the baseline. Redshift surveys are at best comparable if they are
limited to BAO, but they can potentially greatly exceed this if all information
at $k_{\rm max}(z)=
0.1~\left[D\left(z\right)/D\left(0\right)\right]^{-1}\ihmpc$ can be used.

\section{non-Gaussianity}

Many of the non-standard models of how the primordial structure was seeded
predict non-Gaussianity of local type, $\Phi=\phi+\fnl \phi^2$, where $\Phi$
is the gravitational
potential in the matter era and $\phi$ is the corresponding primordial Gaussian
case. The effect of this type of initial conditions on the galaxy power 
spectrum was recently computed by \cite{2008PhRvD..77l3514D} and further 
investigate by \cite{2008ApJ...677L..77M,2008arXiv0806.1061M,
2008arXiv0806.1046A,2008ApJ...684L...1C}. The effect was 
used to place 
observational limits on $\fnl$ by \cite{2008JCAP...08..031S}.
We can easily generalize the \cite{2008arXiv0807.1770S} multi-tracer error on 
the non-Gaussianity parameter $\fnl$ to the  
more transparent approach of this paper.  
In that case, where redshift-space distortions have so far been ignored, the 
perturbations of galaxy type $i$ can be written
\begin{equation}
\delta_{g i}=\left(b_i+c_i \fnl\right) \delta +n_i
\end{equation}
where $c_i$ is an ideally known coefficient characterizing the galaxy type's 
response
to non-Gaussianity (called $\Delta b(k)$ by \cite{2008arXiv0807.1770S}, 
with $\Delta b_i(k)= 3\left(b_i-p_i\right)\delta_c \Omega_m H_0^2/c^2 k^2 
T\left(k\right)D(z)$ from a model for clustering of halos -- see 
\cite{2008arXiv0807.1770S} for an explanation).
The model we will use for perturbations of two types of galaxies is
\begin{equation}
\delta_{g 1}=b\left(1+c_1 \tfnl\right) \delta +n_1~,
\end{equation}
where $\tfnl=\fnl/b$, and
\begin{equation}
\delta_{g 2}=b\left(\alpha+c_2 \tfnl\right) \delta +n_2~.
\end{equation}
Note that the quantity we can measure precisely is not really $\fnl$ but 
instead $\tfnl=\fnl/b$. If one is trying to calculate the detection limit, as 
in \cite{2008arXiv0807.1770S}, 
then the error on $b$ is irrelevant, but for actual constraints on its value 
once it is detected the degeneracy between $b$ and $\fnl$ becomes relevant. 
Like \cite{2008arXiv0807.1770S}, we start by assuming that $\alpha$ is measured 
precisely from relatively small scales where the non-Gaussian effect is 
small, and $c_i$ are known, so we only have two free parameters, 
$P\equiv 2~b^2 \left<\delta^2\right>$, and 
$\tfnl$.
The Fisher 
matrix for a single mode can be inverted to give the error on $\tfnl$, in 
the low noise, small $\tfnl$ limit:
\begin{equation}
\sigma^2_\tfnl=\frac{X_{11} -2 X_{12} + 
X_{22}}{\left(c_1-\alpha^{-1} c_2\right)^2}~, 
\end{equation}
where, again, $X_{ij}=N_{ij}/b_i b_j P_m$.
This result is the same as in \cite{2008arXiv0807.1770S}, once we account for
differences in definitions.

\begin{figure}
\resizebox{\textwidth}{!}{\includegraphics{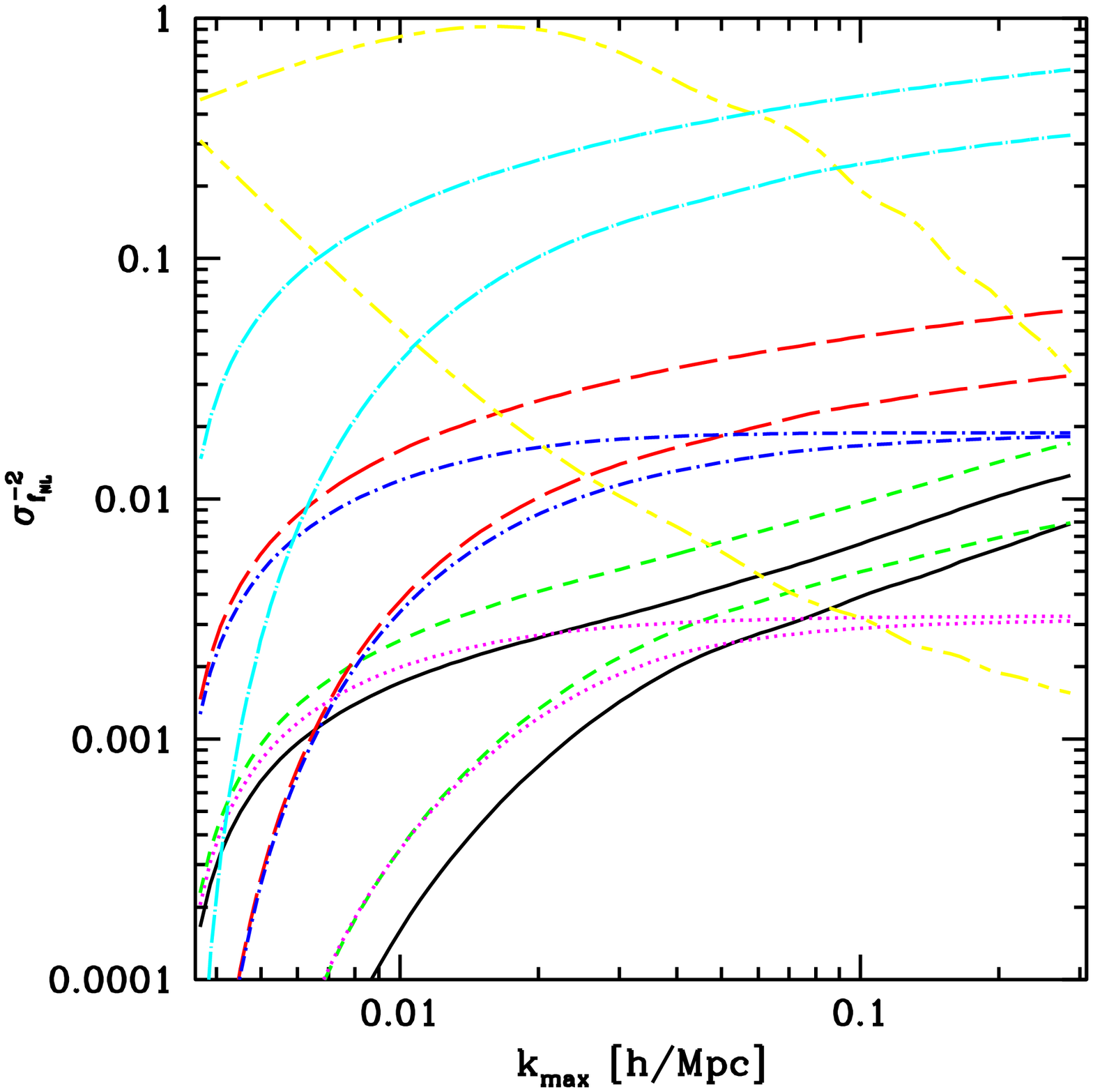}}
\caption{
Projected $\fnl$ errors (inverse variance) as a function of $k_{\rm max}$, for 
the SDSS-III/BOSS galaxy example described in the text. 
Strictly speaking, this is 
not $\sigma_{\fnl}$ but instead $b~\sigma_\tfnl$, where $\tfnl=b^{-1}\fnl$, but
the distinction is irrelevant for a rough detection limit. In 
each case, the lower curve of a type is marginalized over $P$ and $\alpha$, 
while the upper curve is not. Black (solid) assumes BOSS alone, treated
as a single tracer. Green (short-dashed) curves show the 
case when the BOSS galaxies are split into high and low bias subsamples, as
described in the text, while magenta (dotted) curves are the same except the
transfer function is ignored when computing the non-Gaussian effect,
restricting it to large scales. 
Red (long-dashed) show an infinite S/N, unbiased, tracer added to BOSS, while
blue (dot-short-dashed) show the same with no transfer function effect in the 
calculation of the non-Gaussian signal.
Cyan (dot-long-dashed) shows a perfect unbiased tracer 
added to a tracer with $b=1.9$
and ten times the BOSS number density.
The yellow (short-dash-long-dash) curves do not show errors on 
$\fnl$, the upper curve shows $\left(P/2 N\right)/10$,
where $P$ is the power spectrum and $N$ is the noise for the full set of 
BOSS galaxies, and the lower curve $\fnl c\left(k\right)/b$ 
where $c(k)$ is our best guess at the BOSS galaxies' 
response to non-Gaussianity, $b$ is their bias, and $\fnl=30$.
}
\label{figbossfnl}
\end{figure}

\begin{figure}
\resizebox{\textwidth}{!}{\includegraphics{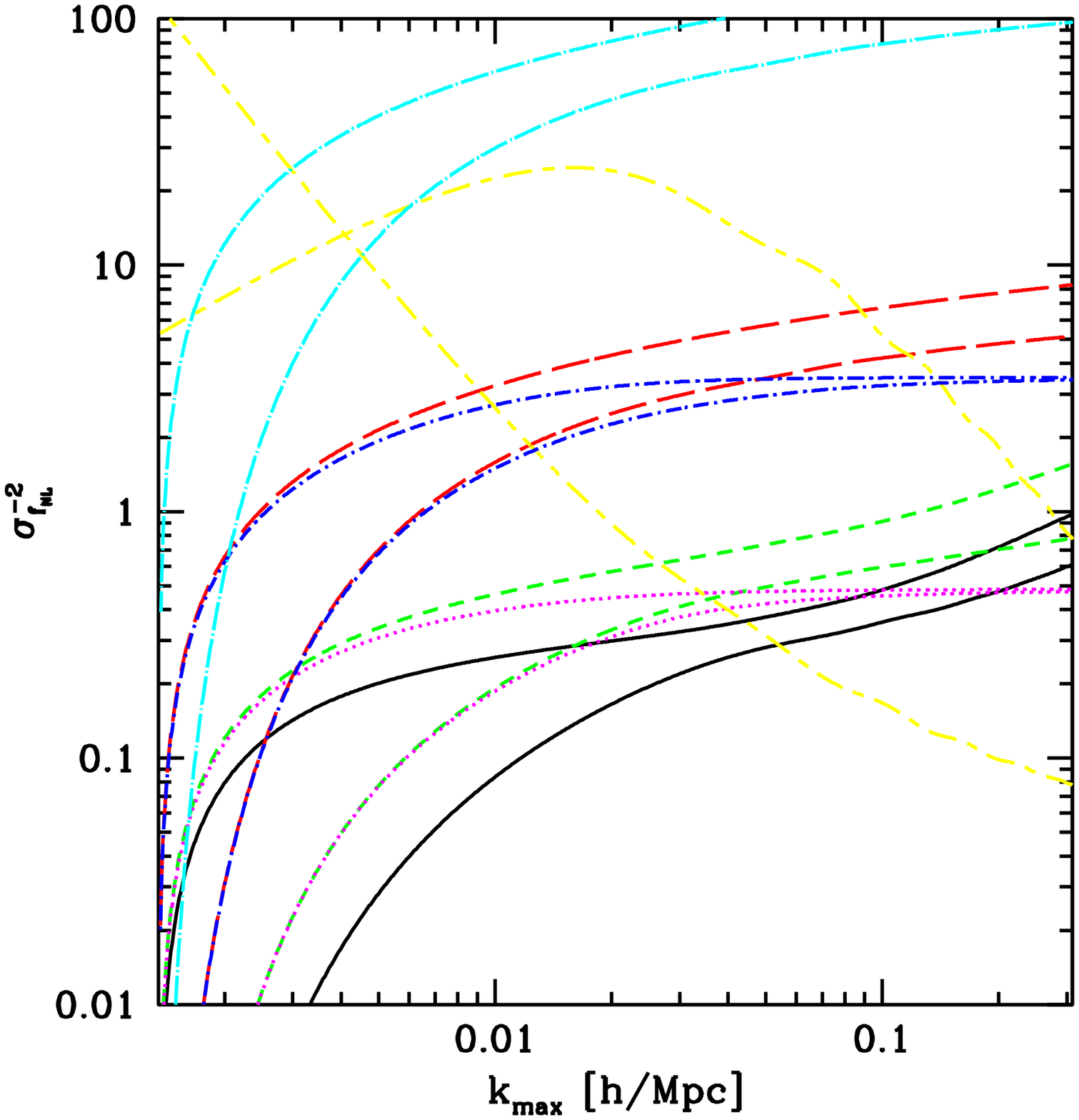}}
\caption{
Projected $\fnl$ errors (inverse variance) 
as a function of $k_{\rm max}$,
similar to Fig. \ref{figbossfnl}, for the EUCLID-like scenario.
In each case, the lower curve of a type is marginalized over $P$ and $\alpha$, 
while the upper curve is not. Black (solid) assumes EUCLID alone, treated
as a single tracer. Green (short-dashed) curves show the 
case when the EUCLID galaxies are split into high and low bias subsamples, as
described in the text, while magenta (dotted) curves are the same except the
transfer function is ignored when computing the non-Gaussian effect,
restricting it to large scales. 
Red (long-dashed) show an infinite S/N, unbiased, tracer added to EUCLID, while
blue (dot-short-dashed) show the same with no transfer function effect in the 
calculation of the non-Gaussian signal.
Cyan (dot-long-dashed) shows a perfect unbiased  tracer 
added to a tracer with $b=2$
and ten times the EUCLID number density.
The yellow (short-dash-long-dash) curves do not show errors on 
$\fnl$, the upper curve shows $\left(P/2 N\right)$,
where $P$ is the power spectrum and $N$ is the noise for the full set of 
EUCLID galaxies, and the lower curve $1000 \fnl c\left(k\right)/b$ 
where $c(k)$ is our best guess at the EUCLID galaxies' 
response to non-Gaussianity, $b$ is their bias, and $\fnl=1$.
}
\label{figeuclidfnl}
\end{figure}

Estimates of the improvement factor for
specific surveys in \cite{2008arXiv0807.1770S} 
are somewhat optimistic, because the improvement
factor was evaluated using the signal-to-noise ratio at $k=0.01 \ihmpc$, 
near the 
peak of the power spectrum. This will give the correct improvement factor for
a measurement using modes on that scale, but, as shown by 
\cite{2008arXiv0806.1061M}, it is 
larger scale modes, where the signal-to-noise ratio for fixed noise will be
lower, that will give the most interesting constraints on $\fnl$. It is also
not obvious when the assumption that $P$ and $\alpha$ are perfectly 
measured using higher $k$ data is safe.
It seems useful to recalculate the constraints in the slightly more general
approach we used for redshift-space distortions, i.e., including $\alpha$ as
a free parameter and integrating over $k$ up to some $k_{\rm max}$.
The results are shown in Figs \ref{figbossfnl} and \ref{figeuclidfnl}. 
Fig. \ref{figbossfnl} shows the same SDSS-III scenarios discussed for 
constraints on $\beta$.  For the single galaxy type case, we expect an 
error $b~\sigma_\tfnl\sim 21$, as long as we can use modes up to 
$k_{\rm max}\sim 0.05 \ihmpc$. Note that the need to marginalize over the 
amplitude
leads to a $\sim 33$\% expansion in the errors.  The improvement when the 
sample is split into high and low bias parts is modest, with 
$b~\sigma_\tfnl\sim 17$.
This is consistent with the finding of \cite{2008arXiv0807.1770S}.
Adding a perfect unbiased tracer helps substantially more, allowing a 
measurement to $b~\sigma_\tfnl\sim 7.3$, again consistent with the 
improvement factor of \cite{2008arXiv0807.1770S}. The marginalization
over nuisance parameters has led to a $\sim 45$\% increase in the error.  

Fig \ref{figeuclidfnl} shows the $\fnl$ constraints for the EUCLID-like 
scenario,  as discussed for $\beta$.  
With only a single tracer, we expect an
error $b \sigma_\tfnl\sim 2.0$, as long as we can use modes up to
$k_{\rm max}\sim 0.05 \ihmpc$.  A simple split of the sample could 
improve this significantly, to 1.4 (note that \cite{2008arXiv0807.1770S} 
assumed a factor of two higher galaxy density for his high redshift example, 
corresponding to the original
SPACE plan, and so found a somewhat larger enhancement factor).
The enhancement when a perfect unbiased tracer
is added is dramatic, $b \sigma_\tfnl\sim 0.54$.  In order to produce a 
solid detection for $\fnl < 1$, we would need either a higher number 
density, e.g., like the SNAP sample we discussed above (except over a larger
volume), or we would need to go 
to higher redshift where there is more volume, 
as discussed by \cite{2008arXiv0806.1061M,2008arXiv0806.1046A, 
2008arXiv0807.1770S}.

\section{Discussion and Conclusions}

Galaxy redshift surveys allow one to measure the 3-dimensional distribution of galaxies, 
which contains an enormous amount of information compared to the 2-dimensional 
surveys 
such as cosmic microwave background anisotropies or weak lensing. Their main limitation 
is that we cannot directly determine the bias $b$ which relates galaxies to the dark matter. 
In this paper we propose a new method to measure the redshift space distortion parameter 
$\beta=f/b$, where $b$ is the tracer bias, 
$f=d\ln D/d \ln a$, $D$ is the growth factor and $a$ the expansion 
factor. The method is based on angular dependence (with respect to the line of sight) 
of relative clustering amplitude of
two tracers with different bias. In such a ratio the dependence on the mode amplitude
vanishes and the method circumvents the usual cosmic variance limit caused by 
random nature of the modes. This allows 
one to determine the relative bias to an accuracy that is only limited by the 
number density of tracers. In principle this allows one to determine the 
velocity divergence power spectrum $\Ptt$ as precisely as one would by directly 
measuring the velocity divergence itself. This, in turn, would allow for a high precision 
determination of growth of structure 
as a function of redshift, as encoded in $fD$. A high density redshift survey 
of galaxies would allow for a more precise determination of dark energy 
parameters, as encoded in DETF Figure of Merit, than any other currently proposed
Stage IV experiment. 

To achieve these gains one needs low noise, which in turn requires
high density of galaxies for both samples. Roughly, the method requires
$\bar{n}P_g \gg 1$, where $\bar{n}$ is the number density and $P_g(k)$ is the 
galaxy power spectrum amplitude. However, most of the existing and planned  BAO
redshift surveys strive for $\bar{n}P_g \sim 1$, since one does not gain much 
in the galaxy power spectrum error once the field is significantly oversampled 
(i.e. $\bar{n}P_g \gg 1$). Most redshift surveys only try to measure 
the shape of the power spectrum (in search of baryonic acoustic oscillations 
and other features) and not its amplitude. 
As long as survey speed is limited by the rate of observing individual objects
(e.g., doubling the number density means halving the volume),
the correct strategy will probably be to observe the full sky at the usual
optimal BAO density, and then to go back to observe at higher density if
possible.  

In addition to very precise measurements of $\beta$, this multiple tracer 
method will allow precise consistency checks of the model. 
As one example, we can measure the bias ratio as a function of scale without 
cosmic variance and therefore study the onset of nonlinear bias. 
As another example, we
can test the redshift space distortion model by fitting for
a parameter $\beta^\prime$ of the distortion model $\left(1+\beta \mu^2+
\beta^\prime \mu^4\right)$, to verify that $\beta^\prime=0$.
This new parameter $\beta^\prime$ can also be measured to the same in principle bottomless level
of precision as $\beta$.

In the short term, the technique in this paper will be most useful
if methods are found that can achieve high effective number density without 
sacrificing volume, e.g., 21cm emission mapping \cite{2008PhRvL.100i1303C}. 
Suppose one had a single perfectly sampled, high resolution field, say, from
21 cm emission mapping \cite{2008PhRvL.100i1303C}.  One can
always create a second, biased field by some simple non-linear transformation, 
like squaring the original field, or taking just the high sigma peaks of the field.  
It may seem impossible that this could gain
one any information, and that would be true if one was able to theoretically
describe and exploit the field on all scales, including using higher order
statistics; however, when one is otherwise limited to only very large scales 
this approach can add information, because it makes selective use of 
small-scale information, exploiting our understanding 
from renormalization of perturbation theory 
that local non-linear transformations of the density field can do nothing but
change the bias parameters on large scales 
\cite{2006PhRvD..74j3512M,2008arXiv0806.1061M}. Another way to see that this
can work is 
simply to note that the two types of galaxy fields we have been discussing are
simply two different non-linear transformations by nature of the same
underlying density field, i.e., there is no reason we cannot do the same thing
artificially.  One would need to keep in mind
in this approach that the transformation is applied to the redshift-space 
field, unlike standard bias which is a transformation of the real-space field. 
For the $\fnl$ application this is unlikely to be a fatal problem, but it may
be for the redshift-space distortion measurement discussed in this paper.

Our results suggest that there is much more to be gained 
by oversampling the density field 
and that one can determine the growth of structure 
as a function of redshift to a higher accuracy than previously believed.  
It is thus worth revisiting the planning of redshift 
surveys in light of these results. 
There is every reason to believe that these measurements can be made very
robustly. We have boiled the constraint from a galaxy survey at a single 
redshift down to one number, but we will still have the full scale and 
angular dependence of the power spectrum, along with higher order statistics,
to help us verify that there is nothing about the modeling that we do not 
understand.  
In addition, a  realistic assessment would also need to address the issue of 
stochasticity between the tracers with numerical simulations, which would 
determine the value of $k_{max}$ beyond which the two tracers are not strongly correlated 
anymore. Preliminary results suggest this is around 
$k_{max} \sim 0.1\ihmpc$ \cite{2004MNRAS.355..129S}, 
but a more
detailed analysis of multi-tracer method with simulations is needed, similar to the recent analysis of 
single tracer method \cite{2008arXiv0808.0003P}.  
We leave this subject to a future work. 

Another promising direction is to combine our no cosmic variance
measurement of $\beta=f/b$ with an analogous measurement
of bias $b$ using a comparison of weak lensing and galaxy clustering 
\cite{2004MNRAS.350.1445P}, to derive $f(z)$ alone without any cosmic 
variance limitation.  We leave these subjects to a future work.

We than Ue-Li Pen, Nikhil Padmanabhan and Martin White for helpful discussions.
U.S. is supported by the
Packard Foundation, DOE and
Swiss National Foundation
under contract 200021-116696/1.

\bibliography{cosmo,cosmo_preprints}

\end{document}